\documentclass{aa}  
\usepackage{txfonts} 
\usepackage{amsmath, amssymb}   
\usepackage{graphicx}      
\usepackage{hyperref}   
\usepackage[dvipsnames]{xcolor}   
\usepackage{booktabs}
\usepackage{lastpage}
\usepackage{microtype}
\usepackage{titlesec}
\usepackage{subcaption}
\usepackage{float}
\usepackage{pifont}
\usepackage{array}

\titlelabel{\thetitle\quad}

\hypersetup{
    colorlinks=true,
    linkcolor=blue, 
    filecolor=magenta,
    urlcolor=cyan,
    citecolor=blue, 
    pdftitle={
Constraining Binary Neutron Star Populations using Short Gamma-Ray Burst Observations
},
    pdfauthor={A. L. De Santis et al.},
}

\newcommand{\ch}{\checkmark}
\newcommand{\x}{\times}

\newcommand{\refsec}[1]{Sec.~\ref{#1}} 
\newcommand{\refeq}[1]{Eq.~\ref{#1}} 
\newcommand{\reffig}[1]{Fig.~\ref{#1}} 

\begin{document}

\title{Constraining Binary Neutron Star Population Synthesis Models using Short Gamma-Ray Burst Data}

\author{
A. L. De Santis
\inst{1,2} 
\and
S. Ronchini
\inst{1,2}
\and
F. Santoliquido
\inst{1,2}
\and
M. Branchesi
\inst{1,2}
      }

\institute{Gran Sasso Science Institute (GSSI), I-67100 L'Aquila, Italy\\
          \email{alessio.desantis@gssi.it}
          \and
          INFN, Laboratori Nazionali del Gran Sasso, I-67100 Assergi, Italy
         }

\abstract
{The landmark multi-messenger observations of the binary neutron star (BNS) merger GW170817 provided firm evidence that such mergers can produce short gamma-ray bursts (sGRBs). However, the limited number of BNS detections by current gravitational-wave (GW) observatories raises the question of whether BNS mergers alone can account for the full observed sGRB population. 
We analyze a comprehensive set of 64 BNS population synthesis models with a Monte Carlo–based framework to reproduce the properties of sGRBs detected by \textit{Fermi-}GBM over the past 16 years. We consider three jet geometry scenarios: a universal structured jet calibrated to GW170817, a universal top-hat jet, and a non-universal top-hat jet with distributions of core opening angles. Our results show that models characterized by low local BNS merger rates ($R_{\text{BNS}}(0) \lesssim 50 \, \text{Gpc}^{-3} \, \text{yr}^{-1}$) predict too few observable sGRBs to reproduce the \textit{Fermi-}GBM population, effectively disfavoring them as sole progenitors. Even when relaxing assumptions on jet geometry, low-rate models remain viable only for wide jets ($\theta_c \geq 15^\circ$), in tension with the narrow jet cores ($\theta_c \approx 6^\circ$) inferred from sGRB afterglow observations. In contrast, models with local merger rates of order $R_{\text{BNS}}(0) \approx 100 \,  \text{Gpc}^{-3} \, \text{yr}^{-1}$ successfully reproduce the observed sGRB population, assuming a plausible fraction of BNS mergers launch relativistic jets and realistic jet geometries. This analysis highlights the power of combining GW observations of BNS mergers with electromagnetic observations of sGRBs to place robust constraints on the BNS merger population and to assess their role as progenitors of sGRBs.
}

\maketitle

\section{Introduction}

\label{sec:intro}

The era of multi-messenger astronomy, ushered in by the joint detection of gravitational waves (GW) and electromagnetic (EM) radiation from the binary neutron star (BNS) merger GW170817 \citep{Abbott2017b, Abbott2017c, Goldstein_2017, Abbot_comment_0, Abbot_comment_1}, has opened unprecedented opportunities to study the physics of compact objects. This event solidified the long-held hypothesis that BNS mergers are the progenitors of short Gamma-Ray Bursts (sGRBs). The connection between BNS mergers and sGRBs, first proposed over three decades ago \citep{Blinkinov1984, Eichler1989}, is now a cornerstone of multi-messenger astronomy, with the model being developed and solidified by subsequent works \citep{Narayan1992, Mochkovitch1993, Nakar2007} before the first direct, unambiguous confirmation from GW170817 and GRB 170817A. However, the robustness and distinctiveness of this connection remains to be quantitatively established. GW observations are opening new possibilities for understanding this relationship, as they now provide constraints on the BNS merger population \citep{LIGOScientificCollaboration2021a,GWTC4}. These constraints can be directly compared with the sGRB sample, informed by more than three decades of observations \citep{vonKienlin2020, Lien_2016}.

A key uncertainty in connecting the BNS and sGRB populations is the efficiency with which BNS mergers produce a sGRB. This efficiency is often encapsulated in the jet fraction, $f_j$: the fraction of mergers that successfully launch a relativistic jet powerful enough to break out of the surrounding ejecta. Previous attempts to constrain $f_j$ have yielded a wide range of values, from a few tens of percent \citep[e.g.][]{Salafia_2022_fj, bhattacharjee_2024} to 100\% \citep[][]{Howell_2019}. The value of $f_j$ is highly uncertain, depending on complex physics including the merger outcome (e.g. long-lived neutron star vs. prompt black hole), the interaction of the jet with merger debris \citep[e.g.][]{Gottlieb2018, Shibata_2019, pavan2025roleinjectionparametersjet}, and the properties of the central engine \citep{Ciolfi2020, Bamber_2024, Hayashi2025}. Similarly, the merger environment itself can shape the angular structure and energetics of an sGRB jet \citep{Pavan_2023}. The jet structure is the main uncertainty when determining $f_j$ as for a given rate of observed sGRBs a small jet fraction can be compensated by a wide angular structure or vice-versa a large fraction with a narrow structure. 

GW170817 represented a critical milestone by demonstrating that BNS mergers are capable of launching relativistic jets, and that these jets possess an intrinsically structured angular profile. The decisive evidence for the presence of a relativistic jet came from Very Long Baseline Interferometry (VLBI), which measured both the apparent displacement of the radio source and its angular size \citep{Moley_2018, Ghirlanda2019}. Multi-wavelength observations of GW170817 over several months provided the first direct observational evidence against a simple top-hat jet model (characterized by a uniform cone of emission) in favor of a structured relativistic jet \citep{, Rossi2002} observed approximately 20 degrees off-axis \citep{Lazzati_2018, Davanzo_2018,Margutti2018, Troja_2018, Lamb2019}. 

Based on the GW170817 observations, population studies often adopt the simplifying assumption of universality, using a structured jet profile calibrated to GRB 170817A \citep[e.g.][]{Ronchini2022}. However, it remains unclear whether the jet structure inferred from GW170817 is representative of the broader sGRB population. Moreover, the assumption of universality itself may not be valid, as variations in merger properties and central-engine physics could naturally give rise to a diversity of relativistic outflows.

In this work, we investigate how gravitational-wave and electromagnetic observations constrain BNS populations, obtained via population synthesis, as progenitors of sGRBs, and how fundamental assumptions about jet physics affect these constraints. We structure our analysis as a comparative study. First, we establish a baseline using a physically motivated, universal structured jet model calibrated to the best fit structure of GW170817 \citep{Ghirlanda2019}. Second, we test how our conclusions change when adopting a universal top-hat jet model. Finally, we explore the effect of relaxing the assumption of universality by allowing jet properties to vary across the population. By comparing the results from these distinct scenarios, we can assess the robustness of our conclusions and identify which properties of the BNS population can be constrained irrespective of the jet physics.

While many studies on sGRB population adopt an empirical approach by convolving the star formation history with a delay-time distribution \citep[e.g.][]{Du2025, pracchia2026shortgammarayburstprogenitors} or by directly parametrizing the merger rate density \citep{Salafia_2023}, our work takes a different path by directly testing the end-to-end predictions from population synthesis models. These models aim at predicting the properties and rates of BNS mergers based on stellar and binary evolution theory \citep[e.g.][]{Dominik2012, Belczynski2018}. Here, we use the comprehensive set of models from \citet{Iorio2023}, which provide distinct merger rate histories based on different assumptions about key evolutionary phases. While many models can produce local rates consistent with GW observations ($R_{\text{BNS}}(0) \approx 7.6 - 250 \, \text{Gpc}^{-3} \, \text{yr}^{-1}$; \citealt{GWTC4}), a critical tension arises when trying to simultaneously explain the observed sGRB rate. Assuming BNS mergers are the sole sGRB progenitors, any model predicting a low intrinsic BNS merger rate can only be reconciled with observations if the jet fraction is correspondingly high. This creates a testable scenario where models requiring an non-physical fraction ($f_j$ larger than 1) can be disfavored. Observed sGRBs carry information beyond their event rate. The full observed properties can be used to constrain the underlying source population and emission physics \citep[e.g.][]{Ghirlanda2016, Ronchini2022}. In this work, we present a systematic framework that accounts for uncertainties in both relativistic jet and BNS population properties, with the goal of connecting theoretical BNS population-synthesis predictions to the observed sGRB properties. We use a Markov Chain Monte Carlo (MCMC) analysis. The code, Multimessenger Analysis with GWs and GRBs in Python (\textsc{MAGGPY})\footnote{\url{https://github.com/LudoDe/gwpop\_LudoDe}}, an optimized implementation based on \citet{Ronchini2022} ($\sim100\x$ speedup), allows us to simulate synthetic sGRB catalogs for each of the 64 analyzed BNS population models. The MCMC sampling adjusts the free parameters of our GRB emission model to find the best fit to the observed \textit{Fermi-}GBM catalog, simultaneously yielding a posterior distribution for the required jet fraction $f_j$. This enables us to assess which populations and BNS merger rates are most plausible, based on their ability to reproduce the observed sGRB population while remaining consistent with GW-inferred local BNS rate constraints.
The paper is structured as follows. In Section~\ref{sec:methods}, we outline our methodology. We first describe the general framework for modeling sGRB emission and introduce the specific jet structure models considered in this work: our fiducial structured jet, the comparative top-hat jet, and non-universal variations. We then present the theoretical BNS population models and the \textit{Fermi-}GBM sGRB observations. Finally, we describe the statistical approach used to construct synthetic sGRB catalogs starting from the BNS population able to reproduce the observed data. Section~\ref{sec:comparison} presents a comparative analysis of the outcomes, directly evaluating how the constraints on BNS populations and jet properties shift depending on the assumed jet model. The results and conclusions are discussed in Section~\ref{sec:discussion} and Section~\ref{sec:conclusions}, respectively. Throughout the article, we use a $\Lambda$CDM cosmology with Hubble constant $\rm H_0 = 67.66\, {\rm km}\, {\rm s}^{-1}\, {\rm Mpc}^{-1}$ and present matter fraction $\rm \Omega_m = 0.31$, from {\em Planck-2018} \citep{Planck:2018vyg}.

\section{Methodology}

\label{sec:methods}

In this section, we detail the framework of our analysis designed to connect BNS population synthesis models with observed sGRB data. 

\subsection{Short GRB Prompt Emission Model}

\label{sec:PEmodel}

To connect a BNS merger population to an observable sGRB population, we model the prompt gamma-ray emission expected from each merger. This model is built upon the intrinsic properties of the central engine and the spectral and temporal evolution of the emission.

\subsubsection*{Intrinsic Engine Properties}

We characterize each burst by three intrinsic, on-axis properties: total radiated energy, characteristic timescale, and rest-frame peak energy. While many studies model the luminosity $(L_{iso})$ function directly  \citep[e.g.][]{Wanderman2015, Salafia_2023, Ghirlanda2016}, we model the distributions of total radiated energy $E_t$ and peak time $t_p$ since these quantities are fundamentally linked ($L_{iso} \sim E_t / t_p$). The total energy converted into radiation, $E_t$, is drawn from a power law with index $k$ and low-energy exponential cutoff $E^*$:
\begin{equation}
P(E_t|k, E^*) = \frac{k}{\Gamma(1-k^{-1})E^*}\left(\frac{E_t}{E^*}\right)^{-k}\exp\left[-\left(\frac{E_t}{E^*}\right)^{-k}\right]
\label{eq:Etdist}
\end{equation}

This form captures the observed distribution for sGRBs while providing a turnover at low energies to ensure the distribution is physically realistic \citep[e.g.][]{Pescalli_2016}.

The peak time of the burst $t_p$ is drawn from a log-normal distribution with median $\mu_t$ and standard deviation $\sigma_t$. Similarly, the on-axis rest-frame peak energy of the $\nu F_\nu$ spectrum, $E_p$, is also drawn from a separate log-normal distribution following \citet{Ronchini2022}. For clarity, both variables $x = t_p, E_p$ are drawn from $\log_{10}(x) = \mathcal{N}(\log_{10}(\mu_x), \sigma_x)$.

\subsubsection*{Spectral and Temporal Properties}

For an observer at redshift $z$ and viewing angle $\theta_v$, the observed peak energy at the peak time is $E_{p, \text{pk}}(\theta_v, z) = R_E(\theta_v) E_p / (1+z)$, and the observed peak time is $\tau_p(z) = t_p (1+z)$. Here, $R_E(\theta_v)$ represents the relativistic Doppler factor \citep[see][]{Ascenzi2020, Ronchini2022}. Although the observed emission duration in MeV energies depends on inclination (see Appendix~\ref{app:figures}), the peak time itself is dominated by the central engine and is angle-independent. We define the instantaneous photon energy spectrum in the observer frame as:
\begin{equation} 
\mathcal{N}(E, t, \theta_v, z) = N_{\text{ph}}(t, \theta_v, z) \cdot f(E, E_p(t, \theta_v, z)) \quad \left[\frac{\text{photons}} {\text{cm}^{2} \text{ s } \text{keV}}\right]
\end{equation}
where $N_{\text{ph}}$ serves as the photon flux normalization as defined in \citet{Ronchini2022}. The term $f(E, E_p)$ describes the time-evolving spectral shape. For this, we adopt a Smoothly Broken Power Law (SBPL) model. While most sGRBs in the \textit{Fermi-}GBM catalog are statistically well-fit by a simpler cutoff power-law (CPL or Compton) model \citep{vonKienlin2020}, this is often a consequence of low signal-to-noise ratio at high energies. Detailed analysis of bright sGRB by \citet{Ravasio_2019} revealed a clear power-law decay above the spectral peak accurately captured by the SBPL form. Additionally, recent magnetohydrodynamic BNS simulations \citep{BNS_modelling} revealed a broken power law spectrum for emitted photons. Based on this evidence, we consider the SBPL model to be a more representative description of the entire sGRB population (see Appendix~\ref{app:figures} for a comparison with other models). The normalized SBPL spectral shape is given by \citet{Ravasio_2018} as:
\begin{equation}
    f(E, E_p) = C_{\text{n}} \left[ \left(\epsilon\frac{E}{E_p}\right)^{-\alpha n} + \left(\epsilon\frac{E}{E_p}\right)^{-\beta_s n} \right]^{-\frac{1}{n}} 
    \label{eq:SBPLshape}
\end{equation}
The term $\epsilon$ is a function of the spectral indices: $\epsilon = \left( - \frac{2 + \alpha}{2 + \beta_s} \right)^{\frac{1}{n(\alpha - \beta_s)}}$. We fix the model's high energy index following \citet{Ravasio_2019}. The low-energy index is instead assumed to be $\alpha = -2/3$. The high-energy index is $\beta_s = -2.6$, and the smoothness parameter is $n=2$. The constant $C_{\text{n}}$ normalizes the spectral shape. We adopt the convention $f(E=E_p)=1$, which sets $C_{\text{n}} = \sqrt[n]{2}$. The temporal evolution of the light curve and peak energy follows the model in \citet{Ronchini2022, Ierardi:2025ibo}. 

From this evolving spectrum, we derive the key observables for comparison with the \textit{Fermi-}GBM catalog. The observed peak photon flux is calculated at the peak time $\tau_p$ and integrated over the GBM band between 50-300\,keV (the most sensitive energy band of NaI detectors)?

\begin{equation}
F_p(\theta_v, z) = \int_{E_1/(1+z)}^{E_2/(1+z)} \mathcal{N}(t = \tau_p, 
E, \theta_v, z)dE \quad \left[\frac{\text{photons}} {\text{cm}^{2} 
\text{ s } }\right]
\label{eq:peak_flux}
\end{equation}

where $E_1 = 50$ keV and $E_2 = 300$ keV. This quantity is additionally calculated on a 64 ms timescale to match the catalog data. The bolometric energy light curve $F(t)$ is given by integrating the energy spectrum at each time step:

\begin{equation}
F(t, \theta_v, z) = \int_0^\infty E \, \mathcal{N}(E, t, \theta_v, z) dE 
\quad \left[\frac{\text{erg}} {\text{cm}^{2} \text{ s } }\right]
\end{equation}

From this, we define the cumulative fluence within the GBM band, $S(t) = \int_0^t \int_{E_1/(1+z)}^{E_2/(1+z)} E  \mathcal{N}(E, t', \theta_v, z) dE dt'$, from which the total fluence and the duration $T_{90}$ are calculated.

\subsection{Predicting the Observed sGRB Rate}
\label{sec:jet_fraction}

We define a fundamental parameter, $f_j$, connecting the BNS merger population to the observed sGRB sample, given by the ratio between the intrinsic sGRB rate ($N_{\text{sGRB}}$) and the BNS merger rate ($N_{\text{BNS}}$)

\begin{equation}
f_j = N_{\text{sGRB}} / N_{\text{BNS}}
\label{eq:fj}
\end{equation}

When $f_j \leq 1$, this quantity can be interpreted as the fraction of binary BNS mergers that are able to produce a successful relativistic jet, namely a jet that is able to break out from the post merger ejecta. In our framework we treat it as a free parameter inferred during the MCMC sampling, which ensures that all populations reproduce the same observed GRB rate. This approach allows $f_j$ to explore values greater than unity as a diagnostic tool to test the viability of the BNS population in reproducing the sGRB data. If the posterior distribution for $f_j$ lies predominantly above $1$, it indicates that the BNS population model under consideration is insufficient to account for the observed sGRB rate.

The predicted rate of sGRBs observed by \textit{Fermi-}GBM is derived by convolving the instrument's duty cycle, the jet production efficiency, the source population properties, and the detector sensitivity. In the most general case the detection efficiency $\Phi$ depends on the redshift, the intrinsic engine parameters $\boldsymbol{\theta}_p$ (e.g. $t_p$ or $E_p$), and the viewing angle $\theta_v$. The predicted rate $R^{\text{pred}}_{\text{sGRB}}$ is

\begin{equation}
    R^{\text{pred}}_{\text{sGRB}} = \epsilon_{\text{GBM}} \cdot f_j \cdot \int \Phi(\boldsymbol{\theta}_p, \theta_v, z) \frac{\mathcal{R}(z)}{1+z} \frac{dV}{dz} \, d\boldsymbol{\theta}_p d\cos(\theta_v) dz
    \label{eq:master_rate_gen}
\end{equation}
where $\epsilon_{\text{GBM}} \approx 0.60$ is the combined effective instrument efficiency and field-of-view \citep{Burns2016}, and $d\cos(\theta_v)$ represents an isotropic distribution of viewing angles. 

For the universal Top-Hat jet model (\refsec{sec:jet_models}), the detection efficiency factorizes because the emission is uniform within a cone of semi-aperture angle $\theta_c$ and negligible outside. In this specific case, an observer detects a burst only if $\theta_v \leq \theta_c$ and the on-axis flux is above the threshold \citep{Matsumoto2019}. This allows us to evaluate the integral of the geometric beaming independently

\begin{equation}
    R^{\text{pred}}_{\text{sGRB}} = \epsilon_{\text{GBM}} \cdot f_j \cdot \underbrace{\int_0^{\theta_c} \sin\theta_v \, d\theta_v}_{\langle f_b \rangle} \cdot \int \Phi(\boldsymbol{\theta}_p, z) \frac{R_{\text{BNS}}(z)}{1+z} \frac{dV}{dz} \, d\boldsymbol{\theta}_p dz,
    \label{eq:master_rate_tophat}
\end{equation}
where $\langle f_b \rangle = (1 - \cos\theta_c)$ is the associated beaming factor. In our non-universal models (\refsec{sec:jet_models}), where $\theta_c$ follows a distribution $p(\theta_c)$, the average beaming factor generalizes to the probability that a random viewing angle is less than the core of a randomly sampled jet

\begin{equation}
    \langle f_b \rangle = \iint  \Theta(\theta_c - \theta_v) \sin\theta_v \, p(\theta_c) \, d\theta_v d\theta_c
    \label{eq:fb_non_universal}
\end{equation}
where $\Theta$ is the Heaviside step function. By requiring $R^{\text{pred}}_{\text{sGRB}}$ to match the observed rate $R^{\text{obs}}_{\text{sGRB}}$, we can infer the necessary jet fraction $f_j$ for each BNS population model for different geometrical assumptions.

\subsection{Jet Structure Models}
\label{sec:jet_models}

The intrinsic energy described in Sec~\ref{sec:PEmodel} is not emitted isotropically but is channeled into a relativistic jet. The geometry of relativistic jets in GRBs is poorly constrained through observations and serves as a major source of uncertainty in population studies. We investigate three distinct assumptions for the jet structure.

\subsubsection*{Universal structured jet}

Our primary model assumes a universal, axisymmetric structured jet, strongly motivated by the afterglow of GRB 170817A. Here, the energy and bulk Lorentz factor decrease with angular distance $\theta$ from the jet's core. We assume angular profiles for the isotropic-equivalent kinetic energy and the bulk Lorentz factor given by:

\begin{align}
\frac{dE}{d\Omega} &= \frac{E_c}{1+(\theta/\theta_c)^{s_1}} \\
\Gamma(\theta) &= 1 + \frac{\Gamma_c-1}{1+(\theta/\theta_c)^{s_2}}
\end{align}
Following \citet{Ronchini2022}, we adopt fixed structural parameters consistent with the best-fit values for GRB 170817A \citep{Ghirlanda2019}: a core opening angle $\theta_c = 3.4^\circ$, an on-axis core Lorentz factor $\Gamma_c = 500$, and power law indices $s_1 = s_2 = 4$. The apparent properties depend on the viewing angle $\theta_v$, with scaling factors for peak energy and flux calculated following the formalism of \citet{Ascenzi2020}. In this framework, the MCMC samples seven free parameters: two for the energy distribution (power law index and cut-off $k, E^*$), two for peak energy and spread($\log_{10}\mu_E, \sigma_E$), two for peak time and spread ($\log_{10}\mu_t, \sigma_t$), and the jet fraction ($f_j$).

\subsubsection*{Universal top-hat jet}

To assess the impact of the assumed jet profile, we also analyze a simplified universal top-hat jet model. As mentioned above this model introduces a degeneracy between the jet fraction and the opening angle given by the efficiency $f_j (1 - \cos(\theta_c)) = const$. This allows us to determine geometries and jet fractions that are compatible with observed jet opening angles. In this framework, we do not use a temporal evolution of the light curve but instead sample the intrinsic peak luminosity $L$ from a distribution analogous to \refeq{eq:Etdist}. The observed peak flux is then calculated as $F_p = L / (4\pi d_L(z)^2 \mathcal{K}(E_p, z))$, where $\mathcal{K}$ is the k-correction factor \citep{Salafia_2023, Poolakkil:2021}. For this model, the MCMC samples six free parameters: two for luminosity power law index and cut-off ($k, L^*$), two for peak energy and spread ($\log_{10}\mu_E, \sigma_E$), the jet fraction ($f_j$) and the core opening angle ($\theta_c$).

\subsubsection*{Exploring Non-Universality}
Finally, we relax the assumption of universality to test whether a diverse jet population significantly alters our constraints on BNS models. In this framework, we extend the top-hat jet model by allowing the core opening angle, $\theta_c$, to vary across the population rather than being a single fixed value. We investigate two distinct scenarios for the population-wide distribution of $\theta_c$, as illustrated in \reffig{fig:model_distributions}. The first scenario is a {\it Flat Model}, where $\theta_c$ is drawn from a uniform distribution between a minimum of $1^\circ$ and a variable upper bound, $\theta_{c}^{\max}$. This model represents a scenario with no preferred physical scale for the opening angle, limited only by the maximum width of the jet geometry. The second scenario is a {\it Log-Normal Model}, characterized by a median angle $\theta_{c}^{\text{med}}$ and a fixed shape parameter $\sigma_{\log_{10}} = 0.5$. This width is chosen to provide a realistic degree of diversity that covers approximately an order of magnitude in core angles, consistent with the heterogeneity observed in sGRB afterglow studies \citep{Rouco_2023}. To ensure physical consistency, both distributions produce minimum angles of $\theta_c^{\text{min}} = 1^\circ$ and specifically for the Log-Normal case we cut angles larger than $\theta_c \geq 45^\circ$. In these non-universal models, the MCMC simultaneously samples the jet fraction $f_j$ and the specific distribution parameters (either $\theta_{c}^{\max}$ or $\theta_{c}^{\text{med}}$). As in the universal top-hat model scenario, this approach allows us to directly quantify the trade-off between the underlying BNS merger rate and the required jet geometry. 

\subsection{\textit{Fermi-}GBM data}
\label{sec:DATA}
GRBs are historically classified into two populations: "short-hard" (sGRBs) and "long-soft" (lGRBs). This bimodality, first identified in \textit{Konus} data \citep{Mazets1981}, was unambiguously established with the \textit{BATSE} catalog \citep{Kouveliotou1993} through the separation of events in the duration-hardness plane.

To calibrate the BNS population with the sGRB observations, we select sGRB data from the \textit{Fermi-}GBM burst catalog\footnote{\url{https://heasarc.gsfc.nasa.gov/W3Browse/fermi/fermigbrst.html}} covering a time span of 16 years (July 2008 - May 2025) (see Appendix\,\ref{app:gbm} for more details). Our approach to select the sample is similar to that of \citet{Ghirlanda2016, Salafia_2023}, aiming to construct a well-defined dataset by applying specific cuts on GRB duration and peak flux. As a proxy for duration, we adopt $T_{90}$, defined as the time interval containing 5\% to 95\% of the total measured fluence. Following established literature \citep[e.g. ][]{Paciesas1999, Lien2016, vonKienlin2020}, we apply the conventional threshold of $T_{90} < 2~\text{s}$ to isolate the sGRB population. Although this threshold may include contamination from short-duration collapsars (single stars) \citep{Zhang2009, Bromberg2013}, we verified that its impact on our population-level results is minimal. In fact, using a more restrictive $T_{90} < 1$~s cut, we find no significant changes in the distribution of GRB observables. Given the inherent difficulty in ensuring progenitor purity solely through duration \citep[see e.g.][]{Ines_2025}, we retain the 2~s limit to maximize the sample size and minimize statistical uncertainty. To mitigate selection effects, we impose a peak photon flux cut of $F_p^{\text{lim}} = 4~\text{ph~cm}^{-2}~\text{s}^{-1}$ (measured on a 64~ms timescale in the 50-300~keV band). We further restrict our sample to events with observed spectral peak energies ($E_p$) between 50~keV and 10~MeV, corresponding to the GBM detection band (see \reffig{fig:data_sel} for both cuts). The peak photon flux $F_p^{\text{lim}}$ cut is chosen to ensure the cumulative peak flux distribution follows the $N(>F) \propto F^{-3/2}$ power law expected for a uniform distribution in Euclidean space \citep{Ghirlanda2016, Ronchini2022}. Notably, while the \textit{Fermi-}GBM catalog provides $E_p$ values derived from Comptonized model fits, our framework employs a Smoothly Broken Power Law (SBPL). Since both models yield consistent $E_p$ values, this discrepancy does not introduce significant bias (see \refsec{sec:PEmodel} for a more thourough discussion). After the cut in peak flux, our sample contains 310 sGRBs. This corresponds to an observed rate of $R_{sGRB}^{obs} = \text{18.61 yr}^{-1}$. This rate is less than the one used in \citet{Ronchini2022}, as a lower value for $F_p^{lim}$ (0.5) was used in that work. After applying the quality cut on the peak energy, we are left with 221 events. For each of these events, we extract the fluence in the 50-300~keV band, $T_{90}$, peak flux in the 50-300~keV band and peak energy. We adopt this band, since the $T_{90}$ of each GRB is computed only in that band in the \textit{Fermi-}GBM catalog. We highlight that even if we impose a cut on the measured peak energy to ensure a good fit quality, that cut is relative only to the GRBs considered for the comparison between the predicted and observed distributions. However, in estimating the true GBM detection rate, we impose no quality cuts and consider all events that triggered the detector, regardless of data quality.

\subsection{Theoretical populations of BNS mergers}
\label{sec:modelvariations}
In this work, we use a comprehensive set of 16 binary population synthesis models from \citet{Iorio2023}. These models explore the large parameter space of stellar and binary evolution uncertainties. Here, we briefly recap how the merger rate density and its redshift evolution are obtained, and highlight the assumptions (see \refsec{sec:commonenvelope} and \refsec{sec:otherassumptions}) that have the most significant impact on the merger rate density.

The models are generated using the \textsc{sevn} code, which evolves a total of $\sim 10^9$ binary systems across 15 metallicity bins spanning the range $Z \in [10^{-4}, 3 \x 10^{-2}]$. The primary masses, $m_{\text{ZAMS,1}}$, are drawn from a Kroupa initial mass function (IMF; \citealt{Kroupa2001}) over the range $[5, 150] \, M_{\odot}$, with a minimum secondary mass of $m_{\text{ZAMS,2}} = 2.2 \, M_{\odot}$.
We use the semi-analytic code \textsc{cosmo$\mathcal{R}$ate} \citep{Santoliquido20, Santoliquido2021} to compute the cosmic merger rates from the merged binaries simulated with \textsc{sevn} \citep{Spera2019, Mapelli2020}. The source frame merger rate density (MRD), $\mathcal{R}(z)$, is computed by convolving the delay-time distribution of the binaries with the cosmic star-formation history:
\begin{equation}
    \mathcal{R}(z) = \int_{z}^{z_{\max}} \int_{Z_{\min}}^{Z_{\max}} \psi(z') \, p(z', Z) \, \mathcal{F}(z', z, Z) \, dZ \, \frac{dt(z')}{dz'} \, dz',
\end{equation}
where $t(z)$ is the look-back time at redshift $z$. The term $\mathcal{F}(z', z, Z)$ is defined as: 
\begin{equation}
    \mathcal{F}(z', z, Z) = \frac{1}{M_{\text{pop}}(Z)} \frac{dN(z', z \mid Z)}{dt(z)},
\end{equation}
where $M_{\text{pop}}(Z)$ is the total initial mass of the simulated stellar population, normalized to the full IMF mass range. The term $\frac{dN(z', z, Z)}{dt(z)}$ represents  the rate of binary compact object mergers forming from stars with initial metallicity $Z$ at redshift $z$ and merging at redshift $z'$, extracted from \textsc{sevn} catalogues. We adopt the analytical fit for the star formation rate density (SFRD), $\psi(z)$, from \citet{Madau2017}:
\begin{equation}
    \psi(z) = a \frac{(1 + z)^b}{1 + [(1 + z)/c]^d} \, [M_{\odot} \, \text{yr}^{-1} \, \text{Mpc}^{-3}],
\end{equation}
with parameters $a = 0.01$, $b = 2.6$, $c = 3.2$, and $d = 6.2$. The metallicity distribution $p(Z \mid z)$ is assumed to follow a log-normal distribution:
\begin{equation}
    p(Z \mid z) = \frac{1}{\sqrt{2\pi\sigma_Z^2}} \exp \left\{ -\frac{[\log(Z/Z_{\odot}) - \langle \log Z(z)/Z_{\odot} \rangle]^2}{2\sigma_Z^2} \right\},
\end{equation}
with $\sigma_Z = 0.1$. From the MRD, we define two key quantities for our analysis, the local BNS merger rate density, $R_{\text{BNS}}(0) \equiv \mathcal{R}(z=0)$, and the total integrated rate, $\Lambda$, defined as:
\begin{equation}
    \Lambda = \int_{0}^{z_{max}} \frac{\mathcal{R}(z)}{1+z} \frac{dV}{dz} \, dz,
\end{equation}
where $\frac{dV}{dz}$ is the differential comoving volume. 

In our analysis, we assume the \texttt{Fiducial} population (\texttt{Model F, $\alpha_{CE} = 1$, $\sigma_z = 1$}) from \citet{Iorio2023} as our baseline. This model yields a local merger rate density of $R_{\text{BNS}}(0) = 112 \, \text{Gpc}^{-3} \, \text{yr}^{-1}$, which lies near the center of the 90\% credible interval reported in the GWTC-4 catalog \citep{GWTC4}. A summary of all analyzed BNS merger populations is provided in Appendix~\ref{app:pop_models}.

\subsubsection{Common Envelope} 
\label{sec:commonenvelope}
The common-envelope (CE) phase is a critical, yet highly uncertain, stage in binary evolution \citep{Webbink1984, Livio1988} and can change predicted BNS merger rates by orders of magnitude \citep{ivanova_common_2013, Broekgaarden22, Santoliquido2021, Vigna_Gomez:2018}. In the CE phase, both stars orbit within a shared and extended envelope. Drag forces remove orbital energy and angular momentum, shrinking the orbit. If the orbital energy deposited into the envelope is sufficient to unbind it, the envelope is ejected, leaving behind a tighter post-CE binary. Otherwise, the stars coalesce. The CE ejection efficiency, $\alpha_{CE}$, sets how effectively orbital energy unbinds the envelope. Low $\alpha_{CE}$ (e.g. 0.5 and 1) requires closer inspirals to eject the envelope, increasing premature mergers during CE while surviving binaries have tighter orbits and shorter GW-driven delay times. \citet{Iorio2023} finds that for  $\alpha_{CE} \leq 1$, premature coalescences suppress BNS formation overall. High $\alpha_{CE}$ (e.g. 3 and 5) favors CE survival at wider separations. Across models, varying $\alpha_{CE}$ shifts the local BNS merger rate density from fewer than 1 to over 300 Gpc$^{-3}$ yr$^{-1}$ (Figs. 19, 20, 22 in \citealt{Iorio2023}), confirming $\alpha_{CE}$ as the dominant parameter in determining the theoretical BNS merger rate \citep{Santoliquido2021}. Within this work we consider the values $\alpha_{CE} = 0.5, 1, 3, 5$.

\subsubsection{Natal Kicks, Mass Transfer Stability and other assumptions}
\label{sec:otherassumptions}

The velocity kicks imparted to neutron stars at birth are important for binary evolution. Models drawing kicks from a Maxwellian distribution \citep[see][]{Hobbs_2005} with dispersion $\sigma=150$ km/s (model \texttt{K150}) yield higher BNS survival and merger rates than models with  $\sigma=265$ km/s (\texttt{K$\sigma$265}), where disruptions are more frequent.

The criteria that determine whether mass transfer from one star to another is stable or instead initiates a common envelope phase can also alter BNS formation channels and rates.  The fiducial model assumes mass transfer is always stable for Main Sequence and Hertzsprung Gap (HG) donors. The \texttt{QCBSE} model instead adopts the standard stability criteria from \citet{Hurley_2002}, widely used in codes like \textsc{MOBSE} \citep{Giacobbo_2018}. The \texttt{QCBB} model goes further, assuming mass transfer is also always stable for pure-Helium star donors which allows for stable mass transfer from pure-Helium stars in scenarios that might otherwise be considered unstable, enhancing the BNS merger rate by providing an additional formation channel, following \citet{Vigna_Gomez:2018}.

Several other physical parameters were tested by \citet{Iorio2023}, but were found to have a comparatively minor direct impact on the overall BNS merger rates. These include variations in Roche lobe Overflow (RLO) accretion efficiency (model \texttt{RBSE}), the specific supernova engine model (\texttt{SND}), the pair-instability supernova model (\texttt{F19}), and the treatment of tides (\texttt{NT}, \texttt{NTC}). While these factors can influence other aspects of compact binary populations or specific BNS properties, their effect on the total BNS merger numbers is secondary to the parameters discussed above.

We analyze sixteen population synthesis models from \cite{Iorio2023}. For each, we consider four values of $\alpha_{CE}$ resulting in a total of 64 BNS merger populations. 

\subsection{Analysis Framework}

\label{sec:mcframework}
From each BNS population, we generate a corresponding synthetic sGRB population calibrated to reproduce the \textit{Fermi-}GBM observations. We employ a hierarchical Bayesian analysis to infer the properties of the sGRB population from the observed data. For each of the three jet structure models described in \refsec{sec:jet_models}, we define a posterior probability distribution for the model's free parameters, $\boldsymbol{\theta}$. We then sample this posterior using a MCMC algorithm, implemented with the \textsc{emcee} package \citep{Foreman_Mackey_2013}. The specific set of parameters $\boldsymbol{\theta}$ and their priors are detailed below for each model.
\subsubsection*{Universal structured jet model}

This model requires a time-evolving light curve and spectrum. The MCMC samples seven free parameters, $\boldsymbol{\theta} = \{k, E^*, \mu_E, \sigma_E, \mu_t, \sigma_t, f_j\}$. Their descriptions and prior ranges are given in Table~\ref{tab:all_priors}.
\begin{table*}[t]
\centering
\caption{Free parameters and prior distributions for the sGRB emission models considered in this work. The common parameters are sampled in every model configuration. The model-specific parameters are included only when the corresponding jet structure is selected. $\mathcal{U}(a, b)$ denotes a uniform prior within $[a, b]$. }
\label{tab:all_priors}
\begin{tabular}{l l c c}
\toprule
\textbf{Parameter} & \textbf{Description} & \textbf{Prior} & \textbf{Models} \\
\midrule
\multicolumn{4}{c}{\textit{Common Parameters}} \\
\midrule
$k$ & Power-law index (Energy or Luminosity) & $\mathcal{U}(1.5, 12)$ & All \\
$\log_{10}(\mu_E/\text{keV})$ & Median spectral peak energy & $\mathcal{U}(2.5, 4.0)$ & All \\
$\sigma_E$ & Standard deviation of $\log_{10}E_p$ & $\mathcal{U}(0.05, 1.0)$ & All \\
$f_j$ & Jet launching fraction & $\mathcal{U}(0, 10)$ & All \\
\midrule
\multicolumn{4}{c}{\textit{Universal structured jet Parameters}} \\
\midrule
$\log_{10}(E^*/\text{erg})$ & Characteristic energy cutoff & $\mathcal{U}(46, 50)$ & Structured \\
$\log_{10}(\mu_t/\text{s})$ & Median peak time & $\mathcal{U}(-2.0, 0.3)$ & Structured \\
$\sigma_t$ & Standard deviation of $\log_{10}t_p$ & $\mathcal{U}(0.05, 1.8)$ & Structured \\
\midrule
\multicolumn{4}{c}{\textit{Top-hat jet Parameters}} \\
\midrule
$\log_{10}(L^*/\text{erg s}^{-1})$ & Characteristic luminosity cutoff & $\mathcal{U}(48, 54)$ & All Top-Hat \\
$\theta_c$ (deg) & Universal core half-opening angle & $\mathcal{U}(1, 25)$ & Univ. Top-Hat \\
$\theta_c^{\text{max}}$ (deg) & Maximum core opening angle ($P_\theta$) & $\mathcal{U}(1, 25)$ & Flat Non-Univ. \\
$\theta_c^{\text{med}}$ (deg) & Median core opening angle ($P_\theta$) & $\mathcal{U}(1, 25)$ & Log-Norm Non-Univ. \\
\bottomrule
\end{tabular}
\end{table*}

\subsubsection*{Universal top-hat jet Model}
In the universal top-hat jet model, the core half-opening angle $\theta_c$ is assumed to be the same for all the sGRBs. The abundance of observed sGRB is given by the efficiency, $\epsilon = f_j (1 - \cos\theta_c)$. This approach allows us to quantify how specific geometric assumptions impact the required BNS abundance ($f_j$). This model does not include time evolution and samples the intrinsic peak luminosity directly. It has six free parameters, $\boldsymbol{\theta} = \{k, L^*, \mu_E, \sigma_E, \theta_c, f_j\}$. While we keep the same prior of $f_j$ as before, we add a flat prior $\theta_c \in [1^\circ, 25^\circ]$. This choice is motivated by observational constraints on sGRB beaming. Population studies of late-time afterglows find a median opening angle of $\langle\theta_c \rangle\approx 6^\circ$ with a tail to wider jets, including measurements and lower limits of $\theta_c\geq 10^\circ$ in $\approx 30\, \%$ of the sample in \citet{Rouco_2023}. Early X-ray afterglow analyses further indicate that most sGRBs are viewed within or very near their jet cores, consistent with compact cores and disfavoring very large core angles \citep{Connor_2024}. Thus, $\theta_c\leq 25^\circ$ is conservative and comfortably includes most of the observed wide core tail without truncating plausible values, while avoiding prior support for unrealistically large cores. All parameters and their priors are are detailed in Table~\ref{tab:all_priors}. 

\subsubsection*{Non-universal top-hat jet models}

The non-universal models expand upon the structured jet and universal top-hat jet frameworks by allowing the core opening angle, $\theta_c$, to vary on an event-by-event basis, according to an underlying population distribution. This approach introduces an additional parameter, $P_\theta$, to the MCMC sampling, which describes the characteristic scale of the core geometry. The sampled parameter set for these models is $\boldsymbol{\theta} = \{k, L^*, \mu_E, \sigma_E, \theta_c, f_j, P_\theta\}$, as detailed in Table~\ref{tab:all_priors}.For the Flat model, $P_\theta$ represents the upper bound $\theta_c^{\text{max}}$. In this scenario, the opening angle is sampled from $\mathcal{U}(1^\circ, \theta_c^{\text{max}})$, effectively testing a population with no preferred scale up to a maximum cutoff ($\theta_c^{\text{max}} <25^\circ)$. For the Log-Normal model, the parameter $P_\theta$ corresponds to the median angle $\theta_c^{\text{med}}$. As noted in the previous section, we fix the dispersion to $\sigma_{\log_{10}} = 0.5$ to ensure the population spans roughly one order of magnitude in jet core angle width. To maintain physical consistency and avoid non-positive or non-physically large angles, the distribution is truncated and renormalized to the range $[1^\circ, 45^\circ]$. In both cases, the simultaneous inference of the jet fraction $f_j$ and the geometric hyperparameter allows us to study the degeneracy between the BNS merger rate and the intrinsic beaming of the sGRB population. We highlight that, although the \textsc{MAGGPY} framework is designed to allow users to define and test arbitrary jet structures, the present analysis focuses exclusively on the three models described above, as they provide a representative coverage of the geometric uncertainties most relevant to the sGRB population.

\subsubsection*{Likelihood}
For each set $\boldsymbol{\theta}$ drawn from the priors, we generate a synthetic catalog of sGRBs. From a population of BNS mergers with redshifts and viewing angles, the sGRB/BNS ratio $f_j$ (see \refeq{eq:fj}) determines the number of successful GRBs. We then draw the intrinsic properties for each event and apply the corresponding jet model to derive the observable quantities. For the structured jet model we derive four key observables for comparison with the \textit{Fermi-}GBM data: fluence ($S_i$), peak flux ($F_{p,64, i}$), peak energy ($E_{p, i}$), and duration ($T_{90, i}$). For the top-hat models, which lacks a temporal profile, the comparison is performed using only the peak flux and peak energy. The final set of selected bursts, $N_{det, sim}$, are those that satisfy the selection criteria from our observational sample. 

The comparison between this simulated catalog and the real \textit{Fermi-}GBM data is based on two components: the observable distribution shape and the event rate. To compare shapes, we use the two-sample Cramér-von Mises (CvM) test, which integrates the squared difference between the Empirical Cumulative Distribution Functions (ECDFs) of the simulated and observed data. This integral approach makes the test sensitive to discrepancies across the entire distribution, rather than a single local maximum. The choice of CvM over the Kolmogorov-Smirnov (KS) statistic used in previous works \citep{Ronchini2022} is motivated by its superior stability and higher statistical power in Monte Carlo sampling \citep{Stephens01091974}. While both statistics are consistent and converge to the same limit, the CvM statistic exhibits lower variance across stochastic simulations \citep{D_Agostino1986}. This provides a p-value, $p_{CvM, X}$, for each observable $X$. To constrain the rate, we use a Poisson likelihood, $P_{Poisson} = \text{Poisson}(N_{det, sim} | \mu_{exp})$, where $\mu_{exp}$ is the expected number of detections. Given the complexity of the detector effects, constructing a direct analytical likelihood is not feasible. Combining these metrics we obtain the likelihood

\begin{equation}
\log \mathcal{L}(\boldsymbol{\theta}) = \sum_{X} \log(p_{CvM, X}) + \log(P_{Poisson})
\label{eq:likelihood_proxy_revised}
\end{equation}
where the sum over $X$ includes the relevant observables. The MCMC maximizes this function to find the sets $\boldsymbol{\theta}$ that best reproduce the observed sGRB population. Details regarding the convergence of our chains and posterior predictive checks are provided in Appendix~\ref{app:mcmc}.

\section{Results}
\label{sec:comparison}

Here, we present the outcomes of our analysis, performed independently for each of the 64 BNS population synthesis models. By comparing the predictions derived under different assumptions for the sGRB jet structure, we assess the physical viability of each BNS model and the robustness of our conclusions. The primary diagnostic is the posterior distribution of the jet fraction $f_j$, which is constrained by the requirement that the model reproduces the observed \textit{Fermi-}GBM sGRB population. By allowing $f_j > 1$, we derive a quantitative metric for model viability which is the degree to which a model fails to produce enough progenitors to match the \textit{Fermi-}GBM rate. 

\subsection{Structured jet population}
\begin{figure*}[t]
    \centering
    \includegraphics[width=\textwidth]{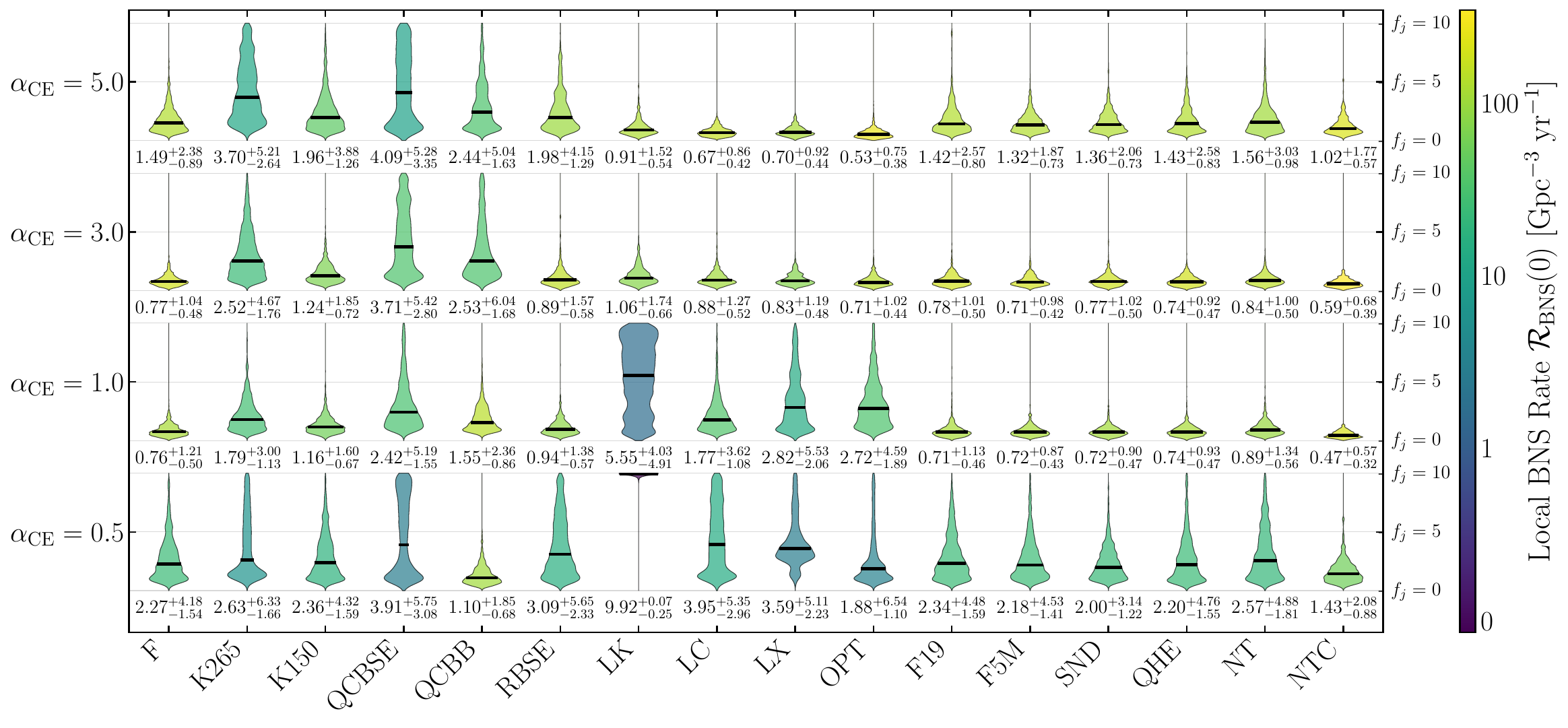}
    \caption{1D marginalized posterior probability distributions of the jet fraction $f_j$ obtained for each considered BNS population synthesis model, assuming a universal structured jet calibrated to GW170817/GRB 170817A. In each plot, the median is indicated by a black line, with its value and 90\% C.I. reported below. The common envelope efficiency $\alpha_{CE}$ is given on the y-axis. The x-axis shows the different models. The color bar indicates the local merger rate of each model. }
    \label{fig:fj_ridge}
\end{figure*}

Applying our analysis to all 64 BNS population synthesis models, and assuming a structured jet as described in \refsec{sec:jet_models}, we obtain the posterior distributions of the jet fraction, $f_j$, shown in \reffig{fig:fj_ridge}.

\begin{figure*}[t]
    \centering
    \includegraphics[width=\textwidth]{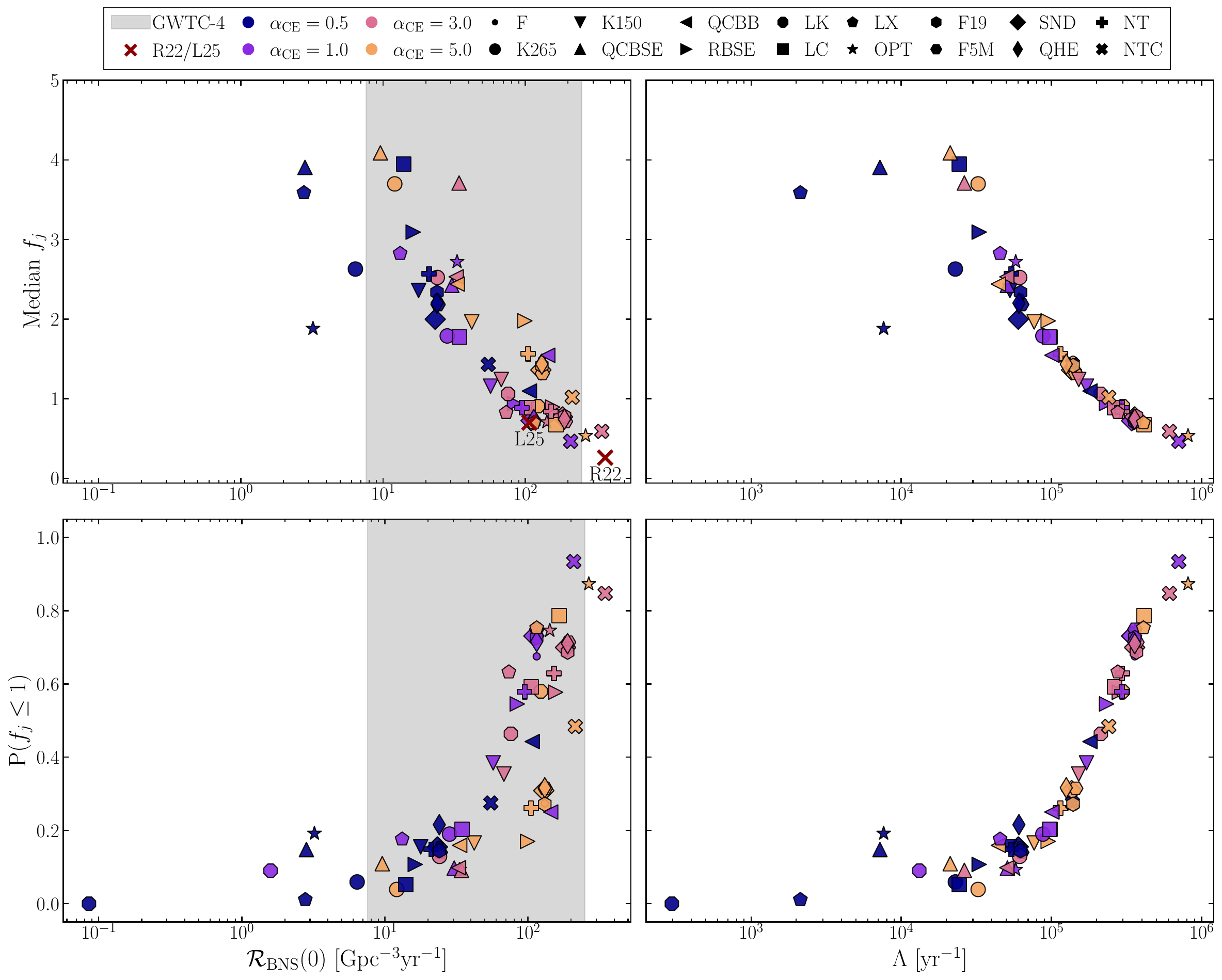} 
    \caption{Jet fraction statistics for the universal structured jet model assuming the GW170817/GRB170817A structure. \textit{Top row:} Median jet fraction $f_j$ versus the predicted local BNS merger rate density $R_\text{BNS}(0)$ (left) and total integrated rate $\Lambda$ (right). \textit{Bottom row:} Quantile of the posterior distribution of $f_j$ at $f_j=1$, as a function of local BNS rate density (left) and total integrated rate $\Lambda$ (right). Symbols are colored according to  the $\alpha_{CE}$ parameter, and each symbol denotes a different population. The shaded vertical region indicates the 90\% credible interval for the BNS merger rate from GWTC-4 \citep{GWTC4}. The cited fractions from the works of \citet{Ronchini2022} and \citet{Loffredo_2025} are given as R22 and L25 for comparison.}
    \label{fig:universal_jet_combined} 
\end{figure*}

\begin{figure*}[t]
    \centering
    \includegraphics[width=\linewidth]{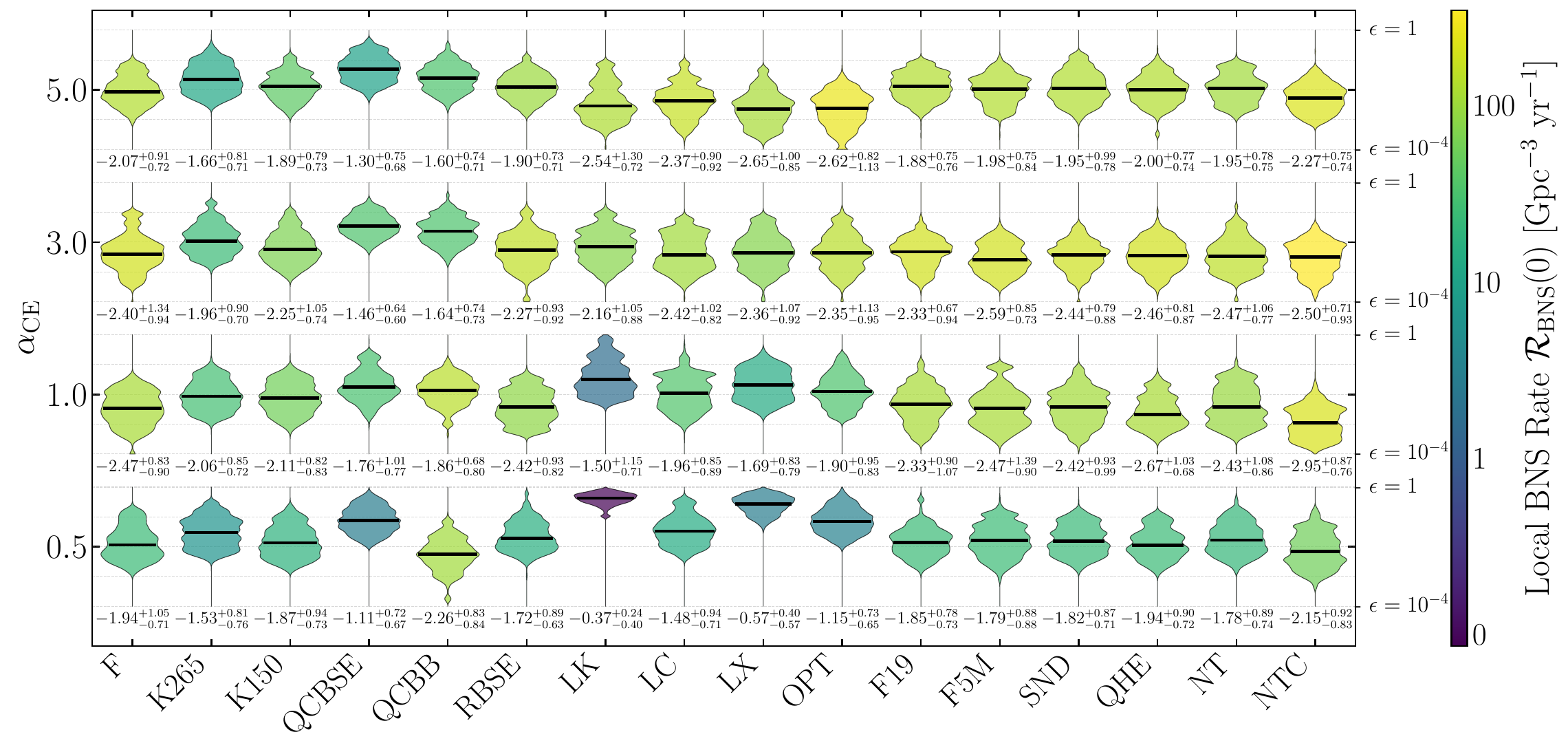}
    \caption{Same as \reffig{fig:fj_ridge}, but with the 1D marginalized distributions of  $\epsilon = f_j (1 - \cos(\theta_c))$ for all the BNS populations, considering the universal top-hat jet model. For clarity $\log(\epsilon)$ is shown.}
    \label{fig:epsilon_posteriors} 
\end{figure*}
\begin{figure*}[t]
    \centering
    \includegraphics[width=\linewidth]{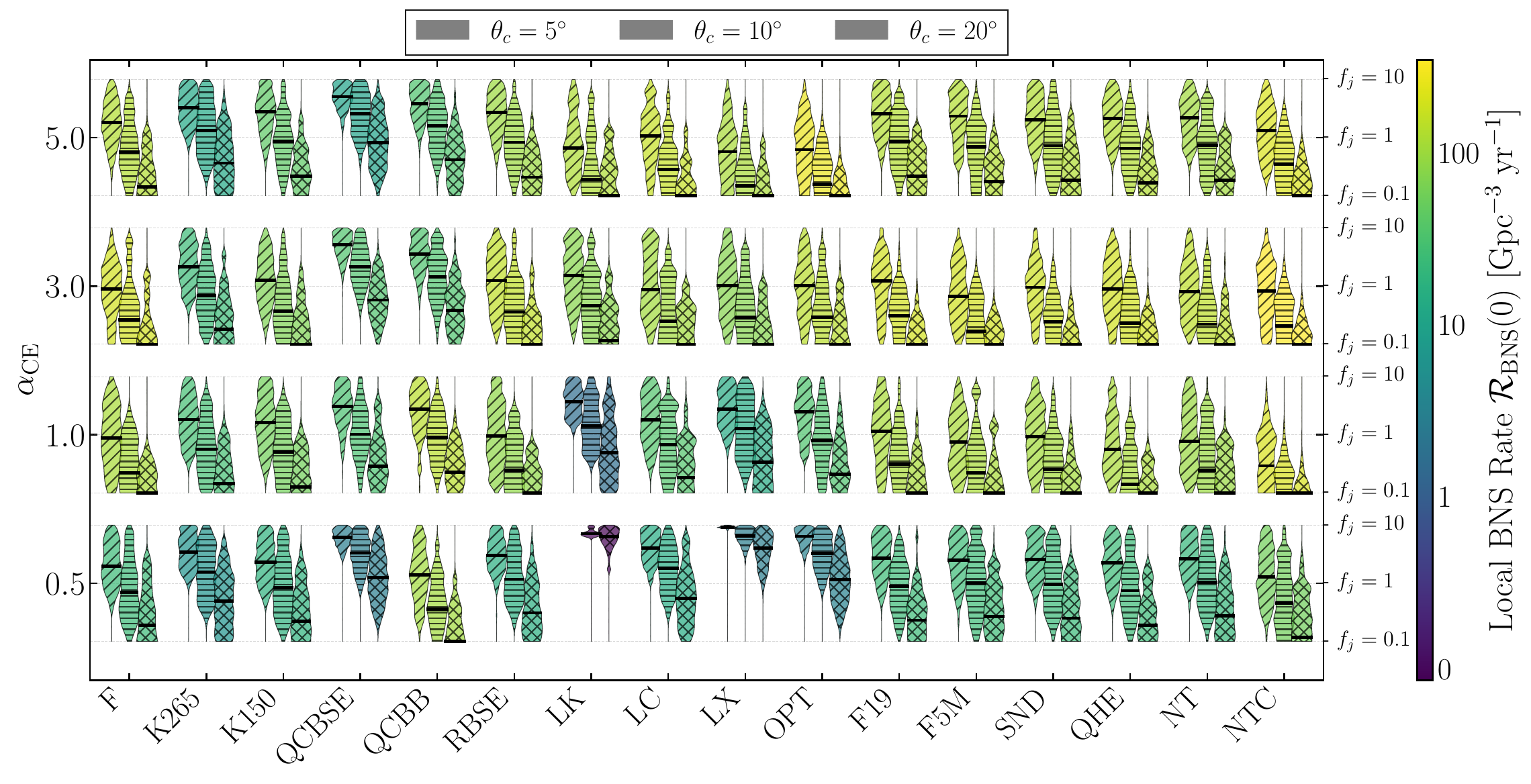}
    \caption{Same as \reffig{fig:fj_ridge} however assuming a universal top-hat jet with three different opening angles $\theta_c = 5^\circ$, $10^\circ$, and $20^\circ$. These posteriors are obtained from the conditioned distribution $P(f_j\mid \theta_c)$ (see \reffig{fig:epsilon_posteriors}). The model \textit{LK} with $\alpha_{CE}=0.5$ does not have any samples below $f_j \leq 10$ for $\theta_c = 5^\circ$.}
    \label{fig:fj_posteriors_5_params}
\end{figure*}
A wide range of $f_j$ posterior distributions is observed, reflecting the significant variation in predicted merger rates across the models spanning orders of magnitude. Models predicting intrinsically high merger rates are compatible with relatively smaller jet fractions. Typically, higher merger rates correspond to larger $\alpha_{CE}$. However, for several models, a trend is observed in which $\alpha_{CE}=5$ yields lower merger rates than $\alpha_{CE}=3$. This behavior is primarily due to the higher efficiency of common-envelope ejection, which leaves the binary system at a wider separation after the common-envelope phase, thereby requiring a longer time to merge. Conversely, for models predicting low intrinsic merger rates (e.g. those with low $\alpha_{CE}$ or high kicks like the model \texttt{K265}) the posterior distributions of $f_j$ are widely spread above the physical threshold of unity. This suggests a scenario for these models where the predicted number of BNS mergers is insufficient to explain the sGRB observations. As detailed in Appendix~\ref{app:boundary_effect}, we allow $ f_j > 1 $ rather than enforcing a strict physical boundary because it provides a quantifiable metric of the missing mergers required by low-rate models, enabling a clearer distinction between marginal and significant model tensions that would otherwise be obscured by saturation at the prior limit. To better understand the relationship between jet fraction and BNS merger rate, we plot the median posterior value of $f_j$ against the local BNS merger, $R_{\text{BNS}}(0)$, and the total event rate obtained by integrating over redshift, $\Lambda$, for each population synthesis model in \reffig{fig:universal_jet_combined} (top row). Here, the anti-correlation between rate and jet fraction is more clearly visible. This correlation exhibits scatter, which is more pronounced when plotting against the local merger rate than the total integrated rate. This scatter arises from models predicting different redshift evolutions, which alters the total number of detectable events in a way that is not captured by the local rate alone.

We also quantify the fraction of the posterior of $f_j$ that lies in the non-physical region by plotting the quantile evaluated at the boundary $f_j = 1$, i.e. $P(f_j \leq 1)$,  shown in the bottom row of \reffig{fig:universal_jet_combined}. Throughout this work, we define a BNS population as physical if the median of its jet fraction posterior satisfies $f_j \leq 1$ (equivalent to requiring $P(f_j \leq 1) \geq 0.5$). These can be considered as the preferred models, i.e., those for which BNS abundance is enough to reconcile with sGRB observations. Notably, we find that models predicting $R_{\text{BNS}}(0) \lesssim 100 \, \text{Gpc}^{-3} \, \text{yr}^{-1}$ exhibit a progressive shift 
of the jet fraction $f_j$ posterior above unity. This indicates that,  under the assumption that BNS mergers are the sole progenitors of sGRBs, explaining the observations would require either an non-physical scenario in which more than one GRB is produced per BNS merger, or the presence of additional progenitor channels.

\begin{figure}[t]
    \centering
    \includegraphics[width=\linewidth]{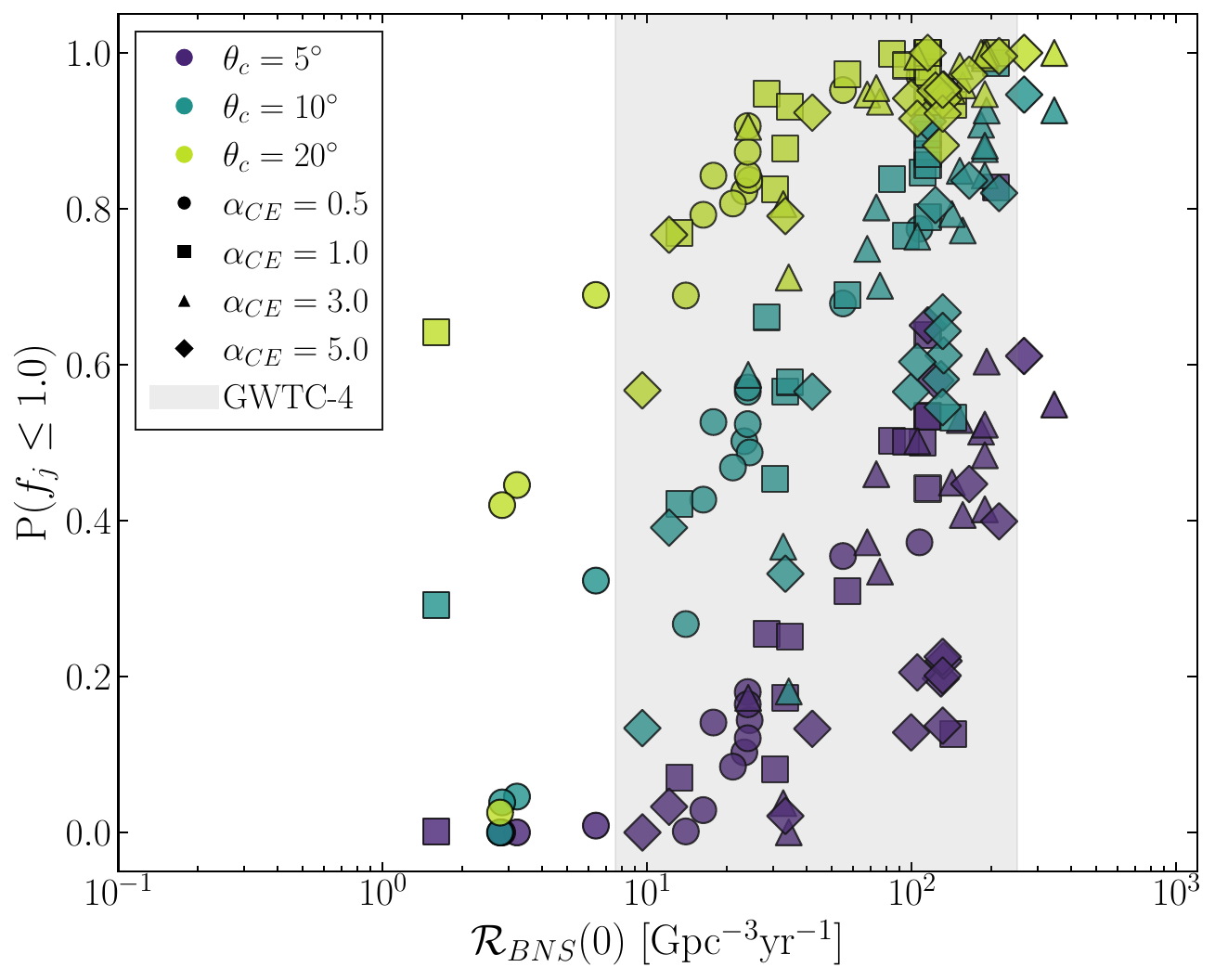}
    \caption{Same as the bottom row of \reffig{fig:universal_jet_combined}, assuming a universal top-hat with three different aperture angles.}   
    \label{fig:quant_3_plots}
\end{figure} 
\begin{figure*}[t]
    \centering
    \begin{subfigure}{0.48\textwidth}
        \centering
        \includegraphics[width=0.8\linewidth]{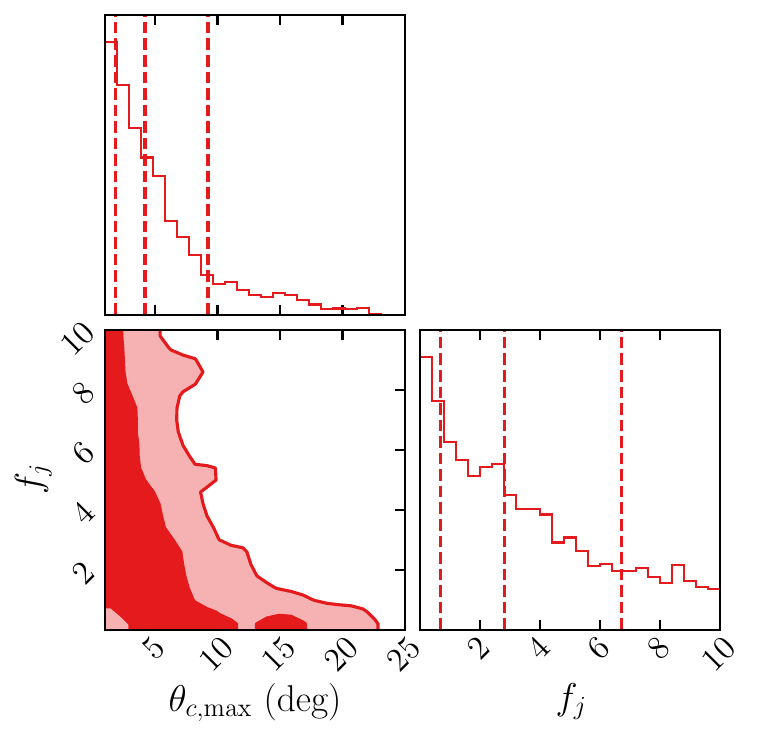}
    \end{subfigure}
    \hfill
    \begin{subfigure}{0.48\textwidth}
        \centering
        \includegraphics[width=0.8\linewidth]{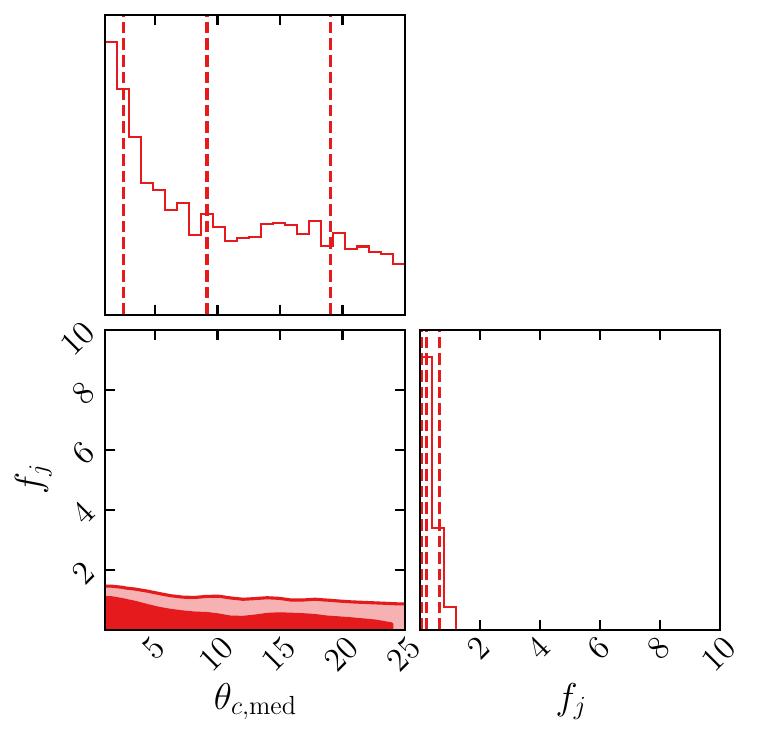}
    \end{subfigure}

    \caption{Two dimensional marginal posterior distribution $P(f_j, \theta_{c,\rm max})$ for the non-universal flat model (left) and $P(f_j, \theta_{c,\rm med})$ for the non-universal log-normal model (right), using our fiducial BNS population. The remaining parameters remain consistent with the results of universal top-hat (see \reffig{fig:simp_model_corner}).}
    \label{fig:non_universal_corners}
\end{figure*}

\begin{figure*}[t]
    \centering
    \includegraphics[width=\textwidth]{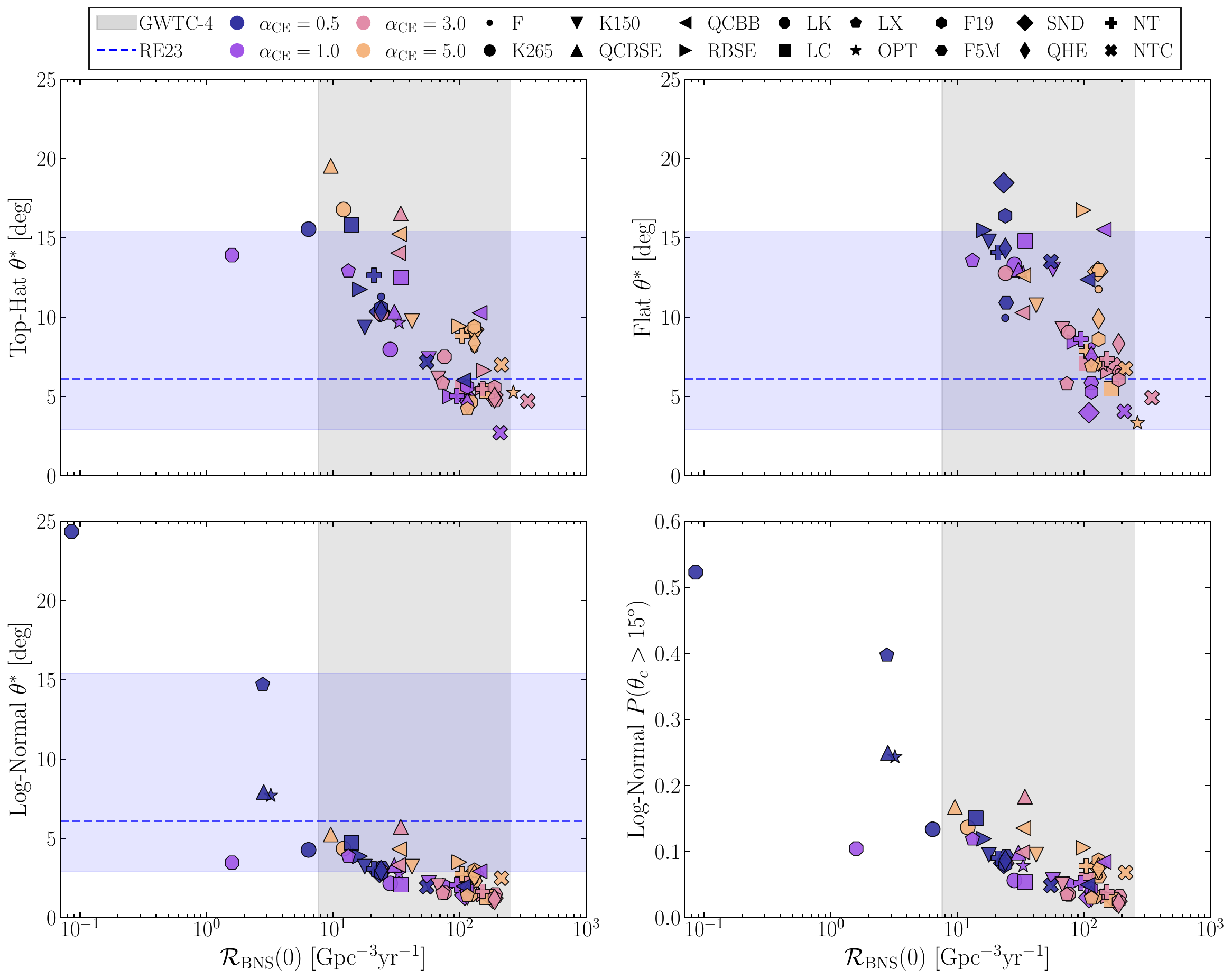}
    \caption{Minimum characteristic jet opening angle, $\theta^*$, required to keep the median jet fraction physical (median $f_j \leq 1$), as a function of the local BNS merger rate $R_{\text{BNS}}(0)$, for the universal top-hat (top left), non-universal flat (top right), and non-universal log-normal (bottom left). Each symbol denotes a different population. The horizontal dashed line and shaded region indicate median and 90\% credible interval of the aperture angle distribution derived from \citet{Rouco_2023}. 
    Bottom right panel shows the fraction of the population with wide jets ($\theta_c > 15^{\circ}$) for the log-normal model. The vertical grey band represents the GWTC-4 90\% C.I. for the BNS merger rate. The plots do not display all 64 analyzed populations, as some require $\theta^*$ larger than the prior upper limit of $25^\circ$. See Table~\ref{tab:viability_grid} for a comprehensive list of all the populations.
    }
    \label{fig:theta_star_combined}
\end{figure*}

\begin{figure*}[t!]
    \centering
    \begin{subfigure}[b]{0.48\textwidth}
        \centering
        \includegraphics[width=\linewidth]{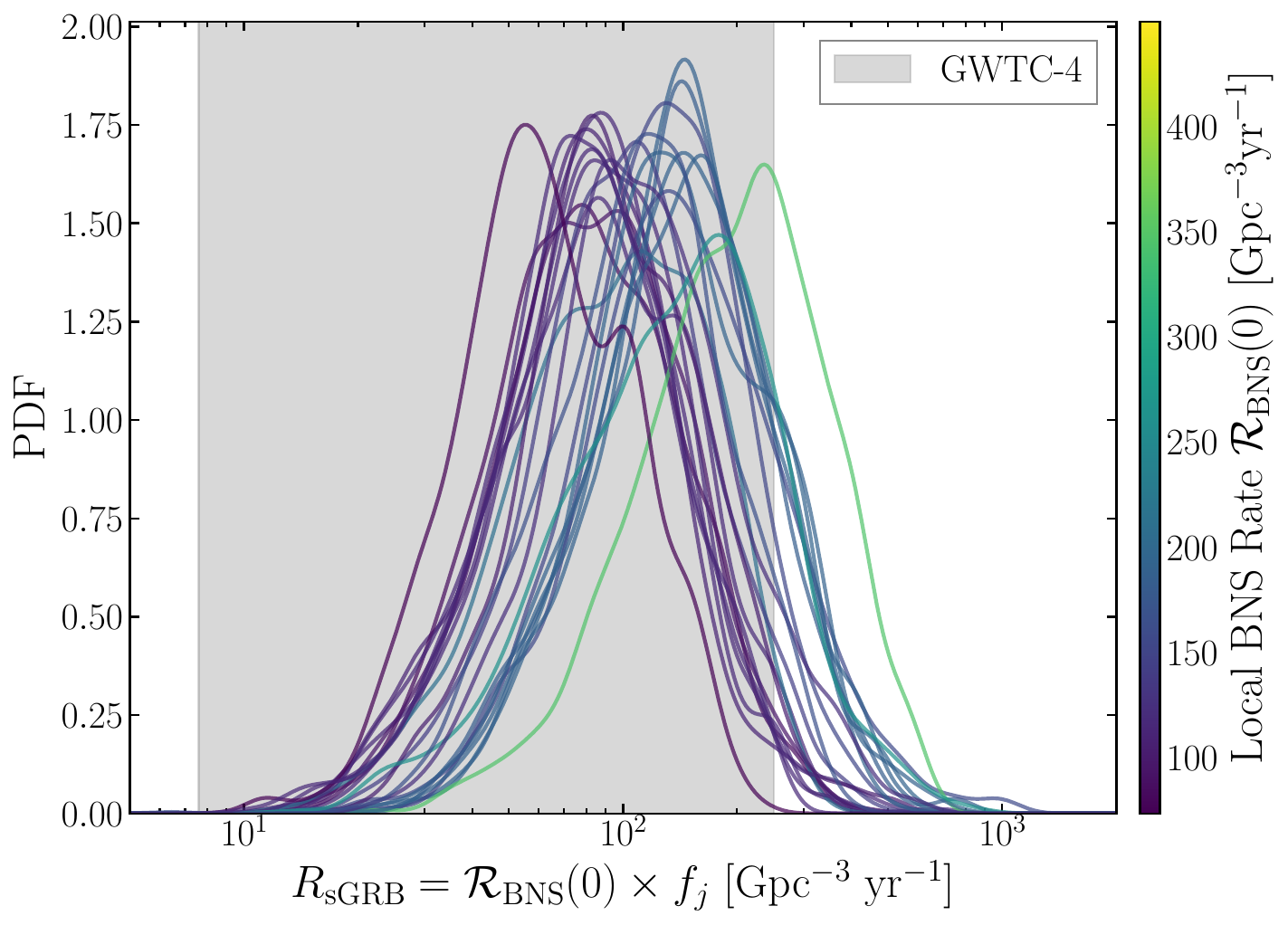} 
        \vspace{-15pt}
        \label{fig:r_sgrb_boxplot_structured}
    \end{subfigure}
    \hfill
    \begin{subfigure}[b]{0.48\textwidth}
        \centering
        \includegraphics[width=\linewidth]{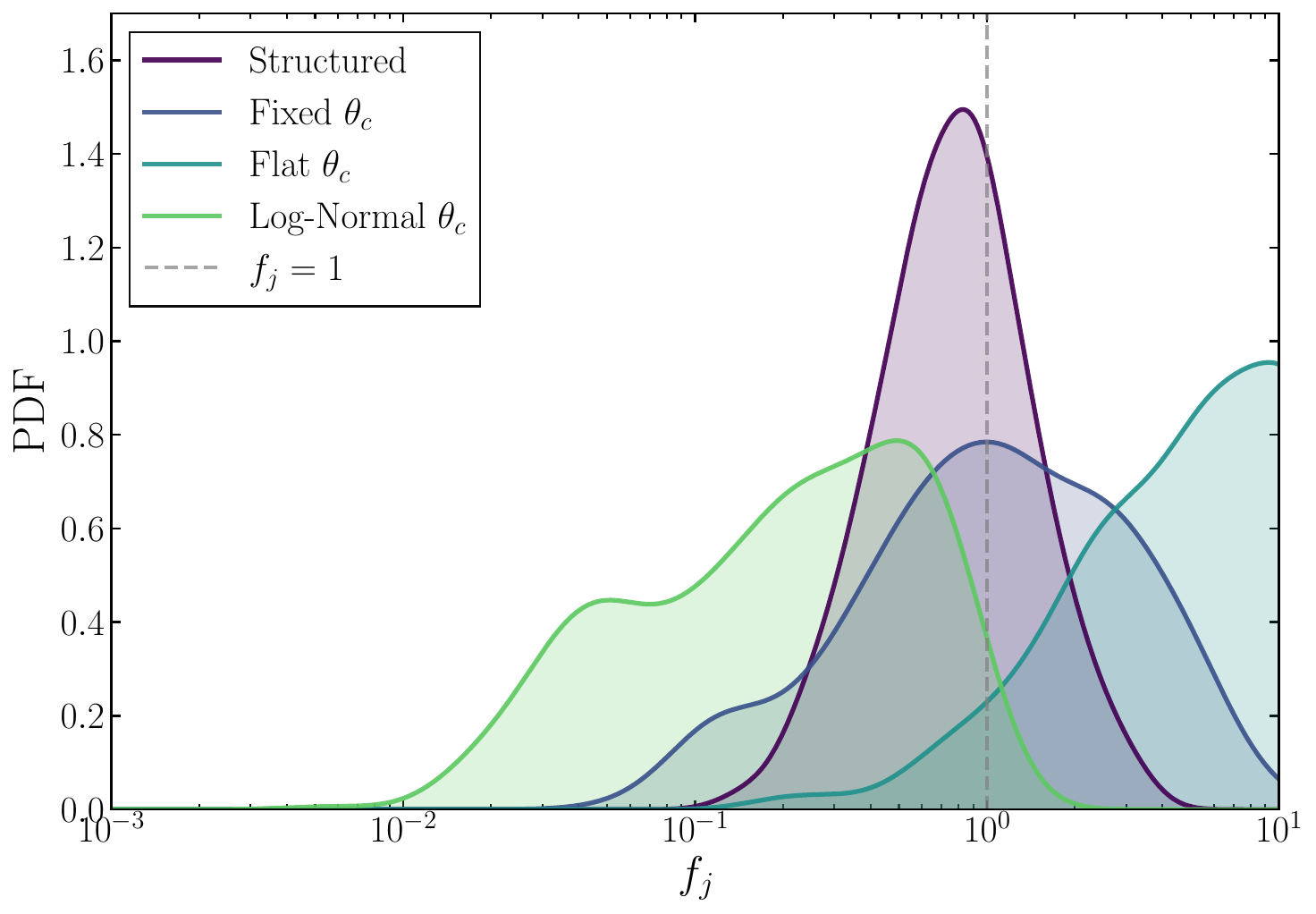} 
        \vspace{-15pt}
        \label{fig:fj_comparison}
    \end{subfigure}
    \vspace{-5pt}
    \caption{\textit{Left:} Inferred intrinsic local sGRB rate $R_{\text{sGRB}}$ for all those BNS population models that allow physical jet fractions (with median $f_j \leq 1$), under the assumption of a universal structured jet. \textit{Right:} Comparison of the inferred $f_j$ posterior distributions for the fiducial BNS population across the four analyzed jet structures: 
    universal structured, universal top-hat, non-universal flat, and non-universal log-normal.}
    \label{fig:combined_rates_fj}
    \vspace{-10pt}
\end{figure*}

\begin{figure*}[t]
    \centering
    \includegraphics[width=\linewidth]{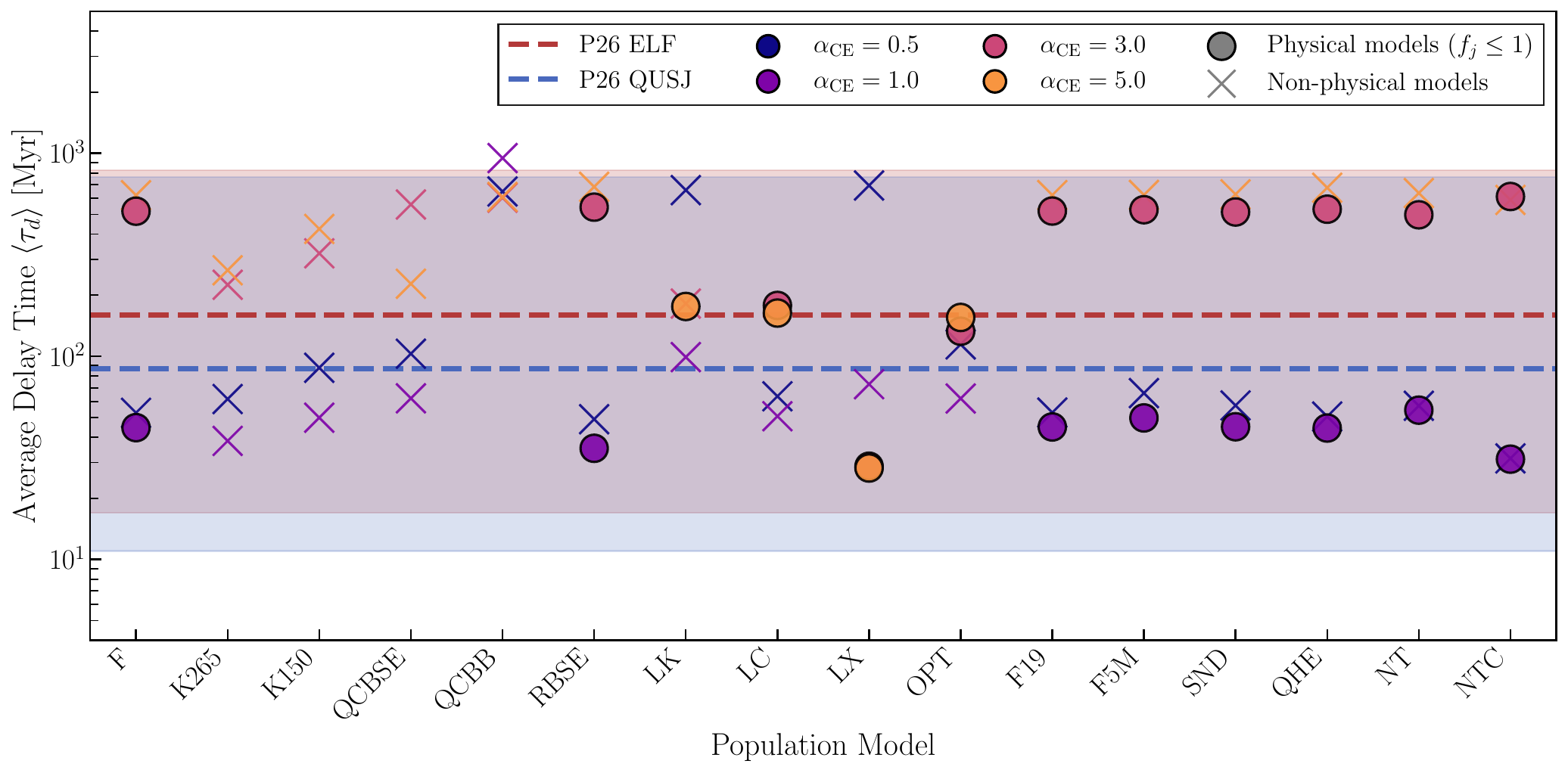} 
    \caption{Average delay time $\langle \tau_d \rangle$ for each of the 64 BNS population models compared to the median and 90\% credible intervals inferred in \citet{pracchia2026shortgammarayburstprogenitors}. The blue and red lines and ranges correspond to two models by \citet{pracchia2026shortgammarayburstprogenitors}: a quasi-universal structured jet (QUSJ) model and an empirical luminosity function (ELF) model, respectively. To highlight the relationship between delay times and model viability, we mark with circles the populations that yield a physical jet fraction (with median $f_j \leq 1$) under the universal structured jet assumption, and crosses those that are considered non-physical.}
    \label{fig:delaytimes}
\end{figure*}

This finding provides a powerful, independent constraint when compared with GW observations. The shaded vertical region in \reffig{fig:universal_jet_combined} shows the 90\% credible interval for the BNS merger rate from GWTC-4 \citep{LIGOScientificCollaboration2021a}. While many of the analyzed models fall within this broad interval, our sGRB analysis effectively establishes a multi-messenger viability window for the BNS mergers as the progenitors of the entire sGRB population.

We can define a conservative lower bound at $R_{\text{BNS}}(0) \lesssim 50  \,\text{Gpc}^{-3}  \text{yr}^{-1}$ for which the majority of the found posterior distribution is within non-physical bounds and for which the MCMC analysis indicates that at at least a factor of 2–3 larger BNS merger rate would be required to reproduce the observed sGRB rate. Conversely, the GWTC-4 bounds $R_{\text{BNS}}(0) \leq 250 \, \text{Gpc}^{-3} \, \text{yr}^{-1}$ give an upper limit on the BNS local rate. 

We note that the range for the BNS local rate from GWTC-4 was obtained by adding only a small portion of the O4 run (almost the first 7 months) to the first three LVK observing runs, and that, to date (Feb 2026), no significant BNS candidate has been released in low-latency in the subsequent one year and half of the O4 run. If the final analysis of O4 data and future GW observations push the allowed parameter space toward BNS merger rates lower than the floor implied by our sGRB constraints, it would become increasingly difficult to account for the full sGRB population with BNS mergers alone. Such a scenario would strengthen the case for a significant contribution from alternative progenitors, such as neutron star-black hole mergers, or that different geometric hypotheses including much wider jet cores are necessary.

We therefore arrive to the conclusion that, under the assumption of a universal GW170817-like jet structure for the sGRB population and that BNS mergers constitute the sole dominant progenitor channel, BNS populations with local merger rates $R_{\text{BNS}}(0) \lesssim 50  \,\text{Gpc}^{-3}  \text{yr}^{-1}$ are difficult to reconcile with sGRB observations.

\subsection{Universal top-hat jet}
\label{sec:results_tophat}
We repeat the analysis using a simplified universal top-hat model. This model constrains the overall efficiency $\epsilon = f_j (1 - \cos(\theta_c))$, which combines the jet fraction and the beaming angle. This allows us to test various values of jet core aperture. In Appendix~\ref{app:mcmc}, we show the posterior distributions for the fiducial BNS population (\reffig{fig:simp_model_corner}). The corresponding median values and confidence intervals are provided in the right-hand columns of Table~\ref{tab:combined_posteriors}. We find that the overall efficiency $\epsilon$ must be low to reproduce the observed data. The posterior distributions of the parameters $k$, $\mu_E$, and $\sigma_E$ closely match those found for the structured model (see Table~\ref{tab:combined_posteriors} and \reffig{fig:corner_plot}). The posterior on $L^*$ is consistent with findings from other works on the GRB luminosity distribution, such as \citet{Wanderman2015} ($\sim10^{51}\,\text{erg/s}$). 

The results for the 64 population models are summarized in \reffig{fig:epsilon_posteriors}. Although the sampler explores $f_j$ and $\theta_c$ separately, we visualize the efficiency $\epsilon$ as the observed rate is directly proportional to this parameter. As expected, models with lower BNS rates require an overall higher efficiency $\epsilon$ to reproduce GRB observables. Owing to the degeneracy between the jet fraction and the jet aperture, this implies that such populations would require jet fractions exceeding unity or unrealistically large jet cores. To visualize this effect and to enable a direct comparison with our structured jet model, we condition the two dimensional posterior of $P(f_j, \theta_c)$ on three fixed values of the opening angle ($\theta_c =$ $5^\circ, 10^\circ, \rm and\ 20^\circ$). \reffig{fig:fj_posteriors_5_params} demonstrates that for both narrow and wide jet assumptions, BNS models with very low intrinsic merger rates yield $f_j$ posteriors that extend significantly above unity.

As in the structured jet case, for each assumed value of the opening angle $\theta_c$, we compute the fraction of the posterior that is physical (ie. $P(f_j \leq 1\mid\theta_c)$) to provide a more direct assessment of which models can reasonably reproduce the observed sGRB rates. A model is considered physical if the median of its jet fraction distribution is $\leq 1$, which corresponds to $P(f_j\leq1)\geq0.5$. The quantiles associated with the 3 analyzed geometries are shown in \reffig{fig:quant_3_plots}, where we observe that for lower opening angles ($\theta_c \approx 5^\circ$), models with local rates $R_{\text{BNS}}(0) \lesssim 50 \, \text{Gpc}^{-3} \, \text{yr}^{-1}$ have almost their entire posterior distribution spread above $f_j > 1$, in agreement with what found in the universal structured jet scenario. To reconcile these low-rate models with observations, significantly larger opening angles ($\theta_c \gtrsim 10^\circ-20^\circ$) are required. This creates tension with observations of sGRB jet breaks, suggesting a low median opening angle $\langle\theta_c\rangle \approx 6^\circ$ \citep{Rouco_2023}. While the top-hat model confirms that the geometric degeneracy allows low rates to be compensated for by wide jets, the tight geometry, inferred for GW170817 and applied to the entire BNS population, implies that BNS merger rates must be relatively high to remain physically viable as the sole progenitors of sGRBs.

\subsection{Non-universality}
\label{sec:results_nonuniversal}
We further relax the assumption of universality by allowing the jet core angle $\theta_c$ in our top-hat jet model to be drawn from a population-wide distribution (uniform and log-normal as defined in Sec\,\ref{sec:jet_models}) for $\theta_c$, treating the distribution's parameters as additional free parameters in the analysis. The posterior distributions from this analysis are shown in \reffig{fig:non_universal_corners} for our fiducial population. We only show the posteriors for the geometry and jet-fraction as the other parameters are almost identical to \reffig{fig:simp_model_corner}. Both models exhibit a similar degeneracy between the parameters $P_\theta$ and $f_j$. One immediate result is that the log-normal model's extended tail allows substantial fraction of the $f_j$ posterior to lie below unity, even for small sampled $\theta_c^{med}$ angles. In contrast, the flat model places a large portion of $f_j$ posterior above unity for almost all small $\theta_c^{max}$ angles. To systematically compare these non-universal results with the universal top-hat model and observational constraints, we introduce a unified geometric metric described below.

\subsection{Narrowest Geometry}
To benchmark all our BNS models against observations, regardless of the assumed top-hat jet model, we quantify here the trade-off between the intrinsic merger rate and the jet geometry. The observed sGRB rate depends on the GRB population beaming factor $\langle f_b \rangle$ and jet fraction $f_j$ such that for a fixed population model (see \refsec{sec:jet_fraction}), the predicted detection rate scales as
$$
R_{obs} \propto f_j \langle f_b (P_{\theta})\rangle
$$
where $P_{\theta}$ represents the parameters of the jet opening angle distribution (e.g. $\theta_c$ for the universal top-hat, $\theta_c^{\text{max}}$ for the flat model, or $\theta_c^{\text{med}}$ for the log-normal model).
Because the observed rate is fixed by \textit{Fermi-}GBM data, a strict degeneracy exists as $f_j \x \langle f_b \rangle = \epsilon$, where $\epsilon$ is a constant specific to each BNS population. If we impose that $f_j$ takes only physical values (median $f_j \leq 1$), there exists a minimum beaming factor $\langle f_b \rangle_{\text{min}}$ required to reproduce the observed event rate. We translate this into a Minimum Characteristic Opening Angle, $\theta^*$, defined as the narrowest population median angle $\tilde{\theta}_c$ that satisfies the physical condition:
\begin{equation}
\label{eq_star}
    \theta^* = \min_{\tilde{\theta}_c} \left\{ \tilde{\theta}_c \mid P(f_j \leq 1) \geq 0.5 \right\}
\end{equation}
We calculate $\theta^*$ for three distinct distribution geometries: 

\begin{enumerate}
    \item \textit{Universal top-hat:} $P(\theta_c) \sim \delta(\theta - \theta_c)$ distribution where $\langle f_b \rangle = 1 - \cos \theta_c$ and the median $\tilde{\theta}_c = \theta_c$.
    \item \textit{Flat distribution:} $P(\theta_c) \sim \mathcal{U}(\theta_c^{\text{min}}, \theta_c^{\text{max}})$, where 
    $$ \langle f_b \rangle = 1 - \frac{\sin \theta_c^{\text{max}} - \sin \theta_c^{\text{min}}}{\theta_c^{\text{max}} - \theta_c^{\text{min}}} $$
    with a population median $\tilde{\theta}_c = (\theta_c^{\text{max}} + \theta_c^{\text{min}})/2$.
    \item \textit{Log-normal distribution:} Defined by a median $\theta_c^{\text{med}}$ and shape $\sigma$, where
    $$ \langle f_b \rangle = \int (1-\cos \theta_c) \, p(\theta_c \mid \theta_c^{\text{med}}, \sigma) \, d\theta_c $$
    with $\tilde{\theta}_c = \theta_c^{\text{med}}$.
\end{enumerate}

For each posterior sample $(f_j, P_\theta)$ we can calculate $\epsilon = f_j \x \langle f_b \rangle$ and derive the median $\langle \epsilon \rangle$. Numerically we solve $\langle f_b(P_\theta) \rangle - \langle \epsilon \rangle = 0$. From the minimum $P_\theta$ we define $\theta^*$ from the associated population median $\tilde{\theta_c}$. An example on how $\theta^*$ is recovered for the universal top-hat model in the case of our fiducial population is shown in \reffig{fig:theta_c_threshold_visual} in Appendix~\ref{app:narrowest}. The figure shows how the $f_j - \theta_c$ posterior follows the predicted line at $\langle \epsilon \rangle = \langle f_j \rangle (1 -  \cos(\theta_c)) = $ const.  We then calculate the $\theta^*$ value for each of the 64 populations across all three geometric assumptions as shown in \reffig{fig:theta_star_combined} and compare these values with the jet opening angles inferred from sGRB afterglow observations \citep{Rouco_2023}. Under the \textit{universal top-hat} assumption (\reffig{fig:theta_star_combined}, top left panel), BNS population models predicting lower intrinsic merger rates require significantly wider $\theta^*$ values to compensate for the lower event count. Models with $R_{\text{BNS}}(0) \approx 100 \, \text{Gpc}^{-3} \, \text{yr}^{-1}$ yield $\theta^*$ values close to the observed median of $\langle \theta_c \rangle \approx 6^\circ$ reported in \citet{Rouco_2023}. In contrast, lower-rate models are forced to assume geometries significantly wider than typical observed values. This trend persists for the non-universal \textit{Flat} model (\reffig{fig:theta_star_combined}, top right panel). As merger rates decrease, the required population median $\theta_c$ moves beyond the $1\sigma$ upper bound of afterglow observations. For some populations, the local rate is too low that no numerical value of $\theta^*$ satisfies the condition of \refeq{eq_star}, since the required value would lie above the maximum value of the prior $\theta_c^{\text{max}}$, effectively excluding these populations as viable progenitors under this assumption. To remain compatible with the narrow geometries inferred from X-ray observations ($\theta_c \leq 10^\circ$), the underlying BNS population must possess a local rate $R_{\text{BNS}}(0) \geq 50 \, \text{Gpc}^{-3} \, \text{yr}^{-1}$. The non-universal \textit{Log-Normal} model (\reffig{fig:theta_star_combined}, bottom left panel) offers more flexibility due to its extended tail, allowing lower-rate BNS models to have a median $f_j \leq 1$ by populating the distribution with wide-angle jets. However, this comes at the cost of physical plausibility. As shown in the bottom right panel of \reffig{fig:theta_star_combined}, for models with $R_{\text{BNS}}(0) \lesssim 20 \, \text{Gpc}^{-3} \, \text{yr}^{-1}$, more than $\sim$10\% of the sGRB population would require core angles $\theta_c \geq 15^\circ$ to maintain a physical jet fraction. This result, under the log-normal assumption, highlights the presence of a subgroup of BNS populations that, in order to have a $f_j$ distribution with a median below one, require a distribution of aperture angles containing a substantial fraction that stands in tension with current observational constraints on sGRB jet collimation.

A comprehensive summary of the physical viability and geometric consistency for all 64 BNS population models across the four analyzed jet scenarios is provided in Table~\ref{tab:viability_grid}. The specific criteria used to evaluate these models, including the treatment of non-physical jet fractions and observational tensions, are further detailed in Appendix~\ref{app:table_details}.

\section{Discussion}
\label{sec:discussion}
Our analysis demonstrates that sGRB observations provide powerful independent constraints on BNS models. By calibrating the sGRB/BNS ratio ($f_j$) to the observed \textit{Fermi-}GBM rate, we identify a clear tension between low-rate progenitor models and electromagnetic observations in terms of sGRB local rate, its evolution, and the inferred jet opening angle.

In the following, we summarize and discuss the results obtained for the various jet models considered.
\begin{itemize} 
\item {Under the assumption of a universal structured jet consistent with GW170817/GRB170817A, BNS population models predicting $ R_{\text{BNS}}(0) \lesssim 50 \, \text{Gpc}^{-3} \, \text{yr}^{-1} $ are effectively ruled out, because the number of BNS mergers is insufficient to account for the full population of observed sGRBs (the infered median is $f_j > 1 $). BNS population models with a local merger rate of $ R_{\text{BNS}}(0) \approx 100 \, \text{Gpc}^{-3} \, \text{yr}^{-1} $, consistent with the current GWTC-4 constraints, are able to reproduce the observed sGRB population provided that the jet fractions are $ f_j \approx 0.7-0.8 $.}
\item {The use of a universal top-hat model allows us to quantify the degeneracy between merger rates and jet geometry. We find that, in order to maintain physically plausible jet fractions, low-rate models are forced to adopt opening angles $ \theta_c \gtrsim 15^\circ-20^\circ $, which are in significant tension with the median value of $ \gtrsim 6^\circ $, inferred from the observation of jet break in afterglow light curves of sGRBs. In contrast, populations with $ R_{\text{BNS}}(0) \approx 100 \, \text{Gpc}^{-3} \, \text{yr}^{-1} $ lie remarkably close to the observed median, allowing for physically plausible jet fractions without requiring wide-angle geometries.} 
\item {This conclusion remains robust when relaxing the assumption of universality by adopting a flat or log-normal distributions for the jet core opening angle. These non-universal jet models show that low-rate BNS population scenarios can reproduce the observed sGRB counts by forcing the jet opening angle distribution toward systematically wider jets. For instance, 
a low–merger-rate BNS population assuming a log-normal jet model requires more than 10-20\% of the systems to have core angles $ \theta_c > 15^\circ $. Such a requirement is not supported by the scarcity of wide-jet observations in sGRB catalogs.}
\end{itemize}

Our finding of physically viable BNS populations capable of reproducing sGRB observations which tend to favor jet-fraction close to unity, is in good agreement with other recent population studies that have taken different approaches to the problem, such as the work by \citet{Salafia_2023}, which utilized a more granular treatment of detector efficiency and inferred local sGRB rates above $ 100 \, \text{Gpc}^{-3} \, \text{yr}^{-1} $, implying a high jet production efficiency. Constraining the jet fraction, has physical implications beyond binary evolution. The ability of a BNS merger remnant to launch a relativistic jet is thought to depend critically on the nature of the central engine \citep{Ciolfi2020}, which is in turn governed by the binary's total mass and the neutron star equation of state (EoS) \citep[e.g.][]{Giacomazzo2013, Ruiz2021}. Our finding of high jet production efficiency disfavors scenarios where prompt collapse to a black hole systematically suppresses jet production, or alternatively implies that even prompt-collapse systems can launch jets efficiently.

We emphasize that the results of our analysis, shown in the different figures, provide a framework for assessing the validity of BNS merger population models in light of the progressively improving observational constraints expected from GW observations and complementary electromagnetic measurements of sGRBs. However, we can already conclude that if future GW observations continue to push the BNS rate toward the lower rates, the discrepancy between the required wide jets and the observed narrow ones \citep{Rouco_2023} would suggest that either BNS mergers are not the sole progenitors of sGRBs, or our understanding of jet observational biases is incomplete.

\subsection{Intrinsic local sGRB rate densities}
\label{sec:literature_comparison}
Using our theoretical and statistical framework, we examine the intrinsic local sGRB rate densities (regardless of the viewing angles) inferred from our models and compare them with values reported in the literature. For each BNS population, we define the rate of successfully launched sGRB jets as $R_{\text{sGRB}} = R_\text{BNS}(0) \x f_j$, where $R_\text{BNS}(0)$ is the local BNS merger rate predicted by population synthesis and $f_j$ is the jet fraction inferred via our analysis. This value represents the all-sky intrinsic local density of sGRB jets produced per year. The resulting distributions for the intrinsic local sGRB rate density, assuming our universal structured jet model, are presented in \reffig{fig:combined_rates_fj} (left panel). We contextualize these results by comparing them to recent derived constraints. \citet{Salafia_2023} (S23) estimated rates using a flux-limited sample similar to ours, finding $R_{\text{sGRB}} \approx 180^{+660}_{-145} \, \text{Gpc}^{-3} \, \text{yr}^{-1}$, while \citet{Rouco_2023} (RE23) reported a rate $R_{\text{sGRB}} \approx 1786^{+6346}_{-1507} \, \text{Gpc}^{-3} \, \text{yr}^{-1}$ ($R_{\text{sGRB}} \approx 361^{+4367}_{-217} \, \text{Gpc}^{-3} \, \text{yr}^{-1}$ for the mock sGRB sample) based on the observed local rate and the jet opening angles determined from the afterglows. We observe that our inferred rates for physically viable populations (specifically those where their median is $f_j \leq 1$) generally sit at the lower end or below the median estimates of S23 and RE23. The higher intrinsic local sGRB rates favored by other studies would increase the tension with the hypothesis that BNS mergers are the sole progenitors of sGRBs, as they would require non-physical jet fractions (median $f_j > 1$). The local sGRB rates are inferred from sGRBs observed out to high redshift, and this tension may also indicate that the redshift distributions assumed in those studies do not accurately reflect the redshift evolution of BNS mergers.

The sensitivity of our conclusions to the assumed jet geometry is illustrated in \reffig{fig:combined_rates_fj} (right panel), where we show the $f_j$ posterior distributions for our fiducial BNS population across the four analyzed jet models. While the universal structured and top-hat models yield comparable constraints, the non-universal models diverge significantly based on the shape of their opening angle distribution. Notably, the log-normal model yields a posterior distribution located almost entirely below $f_j = 1$. This demonstrates that a geometric distribution with a long tail toward wide opening angles can effectively alleviate the tension for lower-rate models. The presence of a sub-population of wide jets increases the average beaming factor, thereby reducing the total number of progenitors required to match the observed \textit{Fermi-}GBM rate. Conversely, the flat model results in a posterior centered well above unity, indicating a much stronger tension. This confirms that for most angles the MCMC is unable to infer physical jet fractions, even BNS models with moderate merger rates may fail to reproduce the observed event counts without non-physical efficiencies. Since $R_{\rm sGRB}\propto f_j$, \reffig{fig:combined_rates_fj} (right panel) can be used to estimate what is the impact of the assumption of jet structure on the inferred sGRB local rate.

\subsection{Delay Time distribution of BNS populations}
Here, we analyze the distributions of delay times between the formation and the merger for BNSs, to compare them with the recent findings of \citet{pracchia2026shortgammarayburstprogenitors} in \reffig{fig:delaytimes}. The delay times are calculated starting from the catalogues of \citet{Iorio2023} and using the code \textsc{cosmo$\mathcal{R}$ate} (see Section\,\ref{sec:modelvariations}). While early studies suggested long delay times (a few Gyrs) for sGRBs, \citet{pracchia2026shortgammarayburstprogenitors} demonstrated that correcting for selection effects in flux-incomplete samples leads to significantly shorter delays, with average values $\langle \tau_d \rangle \approx 10-800$ Myr. \reffig{fig:delaytimes} show the average time delay corresponding to each of our BNS populations. All models but one (QCBB with $\alpha_{\text{CE}} = 1$) exhibit short delay times ($\langle \tau_d \rangle < 1\,\text{Gyr}$) in agreement with the 90\% credible intervals reported by \citet{pracchia2026shortgammarayburstprogenitors}. We also clarify how our time delay average values are derived directly from the BNS population synthesis catalogs and are independent of our sGRB inference. In Appendix\,\ref{app:tdvsz} we also show the redshift evolution of the average delay time.

\subsection{Caveats}
\label{sec:caveats}
\paragraph{Progenitor Purity.} 
A central goal of our analysis is to test the hypothesis that BNS mergers are sufficient to explain the entire observed sGRB population. We use the inferred jet fraction as a diagnostic for this test. Any scenario that requires more SGRBs than BNS progenitors ($f_j > 1$) is interpreted as evidence against the BNS-unique progenitor hypothesis, implying that the BNS population under consideration is not numerous enough to account for the observed sGRB rate. This could suggest a contribution from an alternative channel, such as neutron star-black hole (NSBH) mergers \citep{Colombo2025}. However, the association between NSBH mergers and sGRBs has not yet been firmly established observationally and remains more difficult to quantify. The outcome of NSBH mergers is highly model-dependent and sensitive to the mass ratio, the black hole spin, and the neutron star equation of state \citep{Foucart2012, KrugerFoucart2020}. Numerical simulations and analytical fits suggest that only a subset of the NSBH population, specifically those with high BH spins and/or low mass ratios, yields enough remnant mass to power a jet. These stringent requirements often result in a majority of systems failing to launch a jet \citep{Sarin_2022, Clarke_2025, Biscoveaunu}, implying smaller $f_j \ll 1$ for the NSBH population.
Regarding the population abundance, while some synthesis models predict NSBH rates to be lower than BNS rates by an order of magnitude \citep{Iorio2023}, current observational constraints from GWTC-4 indicate that the two rates may be comparable. Specifically, GWTC-4 reports a BNS merger rate of 7.6-250~Gpc$^{-3}$ yr$^{-1}$ (90\% C.I) and an NSBH merger rate of 9.1-84~Gpc$^{-3}$ yr$^{-1}$.Despite this, because only a small fraction of the NSBH population is expected to produce significant ejected mass and a subsequent jet, we assume that their contribution would not significantly alter the inferred jet fractions presented in this work especially for comparitively higher rate models.
We also acknowledge that the traditional classification distinguishing long and short GRBs has blurred by recent discoveries of long-duration bursts with kilonova counterparts, such as GRB~211211A and GRB~230307A \citep{Rastinejad_2022, Levan_2023}. This suggests that some compact object mergers may be classified as long ($T_{90} > 2~\mathrm{s}$) and thus excluded from our sample. While a detailed treatment of this contamination is beyond the scope of this work, it underscores the need for further observations and improved classification criteria to quantify it. 

\paragraph{Emission Model.} For the structured jet we model the sGRB pulse with a simplified temporal profile (linear rise, power-law decay) and a fixed spectral shape (SBPL with fixed indices). While physically motivated, this does not capture the full diversity and complexity of observed GRB light curves, which often show multiple pulses. This simplification primarily affects the goodness-of-fit for the $T_{90}$ and fluence distributions. While our likelihood is robust to some of this variation, more sophisticated, time-resolved emission models could provide a more detailed fit to the data. Nevertheless, because our analysis is calibrated to observational data, we expect this to have a negligible impact on our conclusions.

\paragraph{Systematic Uncertainties.} The analysis is subject to systematic uncertainties stemming from observational data and instrumental effects. The \textit{Fermi-}GBM detection efficiency and sky exposure ($\epsilon_{GBM}$) are approximations, and the measurement errors for cataloged observables like $E_p$ are not fully propagated in our comparison. Given the large quantity of data spanning over a decade of measurements this is secondary to the major uncertainties in population synthesis and jet modeling. Our predictions of the local sGRB are compatible with \citet{Salafia_2023}, where the authors performed a similar analysis leveraging a more sophisticated treatment of the detector's response. Even with these additional considerations they infer local sGRB rates that are in tension with the bounds from GWTC-4, showing that lower BNS merger rates are disfavored regardless of the level of refinement in instrument response modeling.

\section{Conclusions}
\label{sec:conclusions}

In this work, we present a comprehensive multi-messenger framework connecting 64 state-of-the-art BNS population synthesis models to the 16-year sGRB archive from \textit{Fermi-}GBM. By treating the jet launching fraction, $f_j$, as a free parameter inferred directly from observations, we systematically evaluate the viability of BNS populations under three distinct jet structure assumptions: a universal structured jet calibrated to GW170817, a universal top-hat jet, and non-universal populations with diverse opening angles.

Our analysis identifies a significant tension between BNS population models predicting low local merger rates ($R_{\text{BNS}}(0) \lesssim 50 \, \text{Gpc}^{-3} \, \text{yr}^{-1}$) and the observed sGRB event count. Under the assumption of a GW170817-like structured jet, these models consistently require non-physical jet fractions (median $f_j > 1$) to match observations, implying that the underlying BNS population is insufficient to serve as the sole progenitor of sGRBs. This effectively establishes a lower bound for the merger rate derived purely from electromagnetic observations.

We further demonstrate that this tension cannot be resolved by simply invoking different jet geometries. Using both universal and non-universal top-hat models, we found that low-rate populations can only reproduce the observed sGRB rate if the population is dominated by wide jets ($\theta_c \gtrsim 15^\circ$). This requirement is very difficult to reconcile with the narrow core angles ($\langle \theta_c \rangle \approx 6^\circ$) inferred from sGRB afterglows and constraints from GRB 170817A \citep{Ghirlanda2016, Rouco_2023}. Even when allowing for a log-normal distribution of opening angles, low-rate models are forced to populate the distribution's tail with a fraction of wide jets that is incompatible with current sGRB observational limits.

Consequently, our analysis favors BNS models with local merger rates of $R_{\text{BNS}}(0)$  around $100 \, \text{Gpc}^{-3} \, \text{yr}^{-1}$. In the context of binary evolution parameters, this preference points toward models assuming standard or high common envelope efficiencies ($\alpha_{\text{CE}} \geq 1$) and moderate natal kicks, which are necessary to maintain a sufficient population of merging binaries. Conversely, models characterized by low ejection efficiencies ($\alpha_{\text{CE}} \leq 0.5$) or high natal kicks (e.g., $\sigma = 265$ km/s) are disfavored, as they suppress the merger rate below the one required by electromagnetic observations (see Appendix\,\ref{app:table_details} and Appendix\,\ref{app:pop_models} for specific models). The favored populations naturally reproduce the spectral and temporal properties of \textit{Fermi-}GBM sGRBs with physically plausible jet fractions ($f_j \approx 0.8$) and opening angle distributions consistent with sGRB afterglow observations. The inference of relatively high jet production efficiencies suggests that the conditions for launching a relativistic jet are satisfied in the majority of BNS mergers, disfavoring scenarios where prompt collapse to a black hole systematically suppresses jet formation.

In conclusion, this study highlights the importance of combining GW and electromagnetic observations to constrain the physics of binary evolution. While current GW measurements primarily set an upper bound on the BNS merger rate, sGRB population statistics provide a complementary lower bound. As future GW observing runs further tighten the constraints on the local BNS rate density, the allowed parameter space will progressively shrink, enabling the relation between the jet fraction and the merger rate to become a precise probe of the physical conditions governing binary neutron star mergers.

\begin{acknowledgements}
The authors thank Gor Oganesyan, Biswajit Banerjee, and Matteo Pracchia for insightful discussions and constructive comments that helped improve this manuscript. M.B. and S.R. acknowledge support from the Astrophysics Center for Multi-messenger Studies in Europe (ACME), funded under the European Union’s Horizon Europe Research and Innovation Program, Grant Agreement No. 101131928. F. S. has been funded by the European Union – NextGenerationEU under the Italian Ministry of University and Research (MUR) - CUP D13C25000700001. F.S. has been funded by the European Union –NextGenerationEU under the Italian Ministry of University and Research (MUR) "Decreto per l’assunzione di ricercatori internazionali post-dottorato PNRR" - Missione 4 "Istruzione e Ricerca" Componente 2 "Dalla Ricerca all’Impresa" del PNRR - Investimento 1.2 “Finanziamento di progetti presentati da giovani ricercatori” - CUP D13C25000700001. 
\end{acknowledgements}

\bibliography{references} 

\appendix
\section{GRB modelling}
\label{app:figures}
\begin{figure}[h!]
    \centering
    \includegraphics[width=0.9\linewidth]{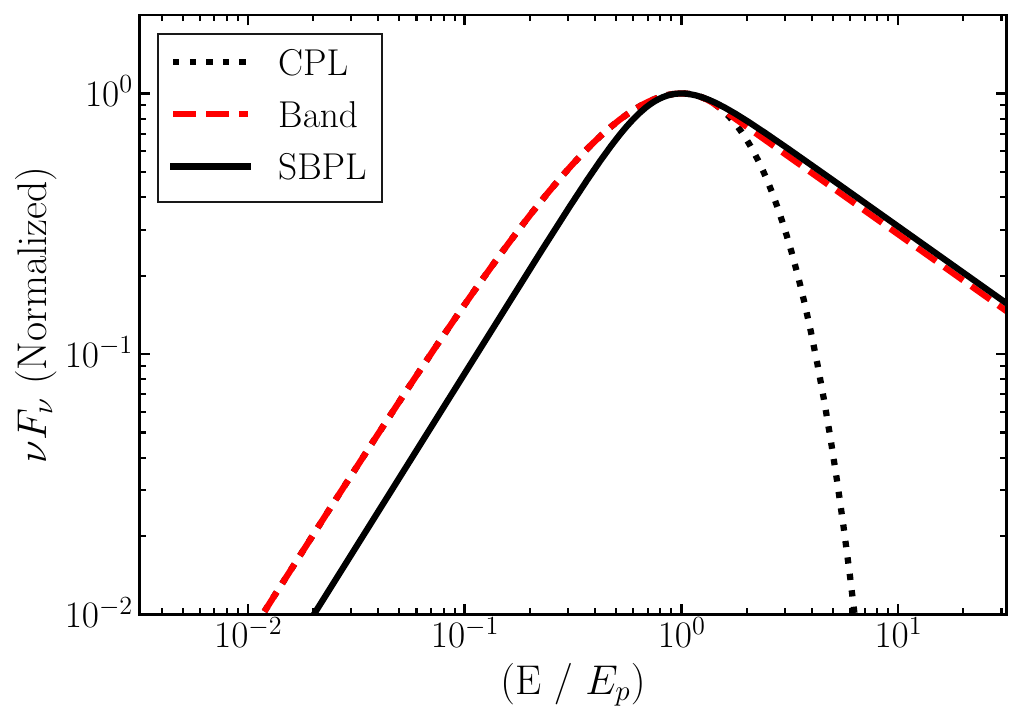}
    \caption{Comparison between different spectral models, normalized to 1 at their maxima. The low and high energy index used is $\alpha = -2/3$ and $\beta = -2.59$ for all models.}
    \label{fig:ModelComparison}
\end{figure}

\begin{figure}[h!]
    \centering
    \includegraphics[width=0.9\linewidth]{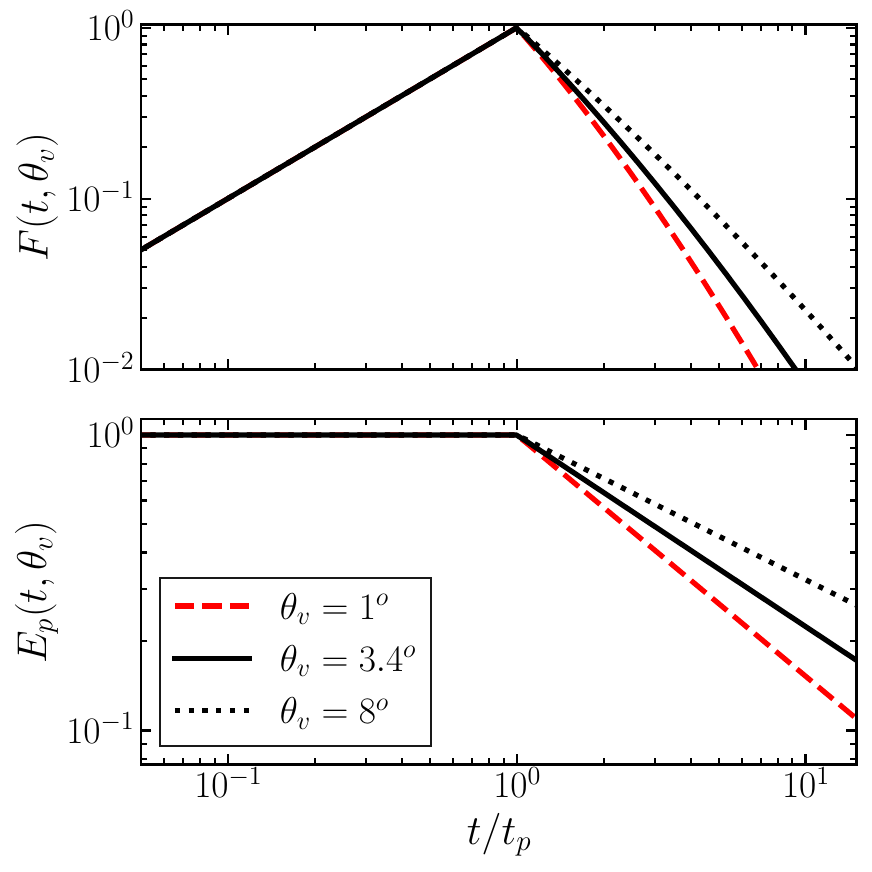}
    \caption{Qualitative behavior of the rest frame peak energy and light curve used in our model with respect to peak time.}
    \label{fig:qualitative}
\end{figure}

\begin{figure}[t!]
    \centering
    \includegraphics[width=0.48\textwidth]{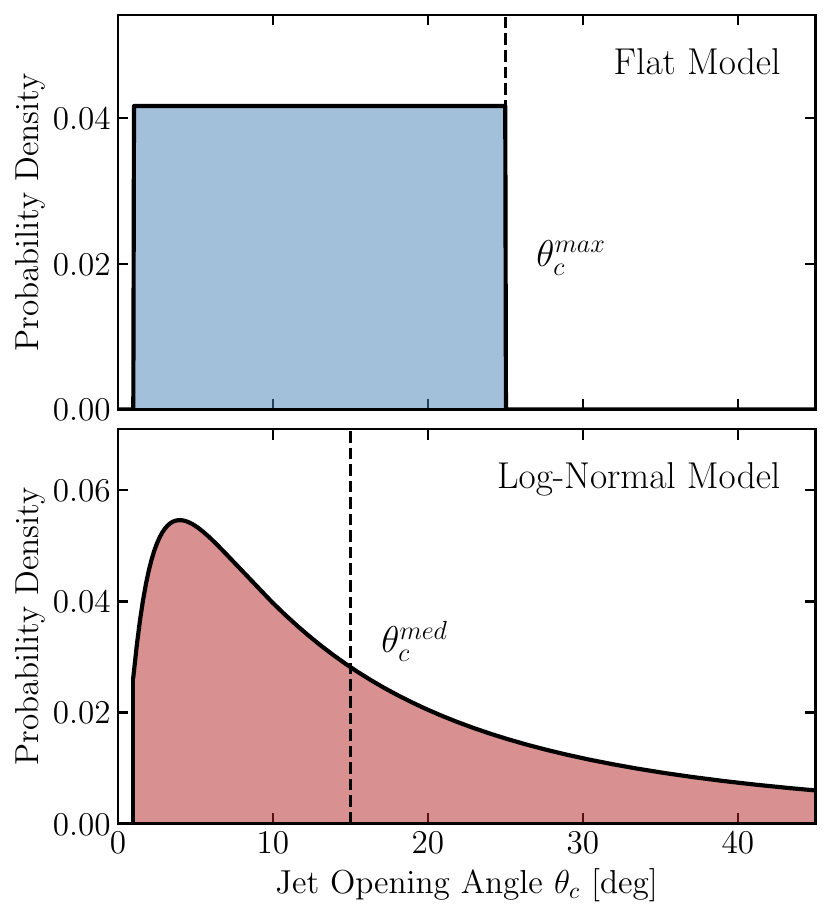}
    \caption{Example distributions for the two non-universal population models in our analysis.} 
    \label{fig:model_distributions}
\end{figure}

In this appendix, we provide supplementary details regarding the sGRB emission model and the statistical validation of our results. \reffig{fig:ModelComparison} compares different spectral models, highlighting the Smoothly Broken Power Law (SBPL) used in this work, which avoids the exponential cutoff of the Comptonized model while providing a smoother transition than the Band function. \reffig{fig:qualitative} depicts the qualitative time evolution of the peak energy and light curve for different viewing angles, showing how inclination primarily affects the post-peak decay slope. Finally, regarding the non-universal jet models discussed in \refsec{sec:results_nonuniversal}, we visualize the probability density functions used for the jet core opening angle $\theta_c$ in \reffig{fig:model_distributions}. The analysis samples the characteristic parameter ($P_\theta$) for these distributions (either $\theta_c^{\max}$ or $\theta_c^{\text{med}}$).

\section{\textit{Fermi-}GBM data}
\label{app:gbm}
\begin{figure}
    \centering
    \includegraphics[width=\linewidth]{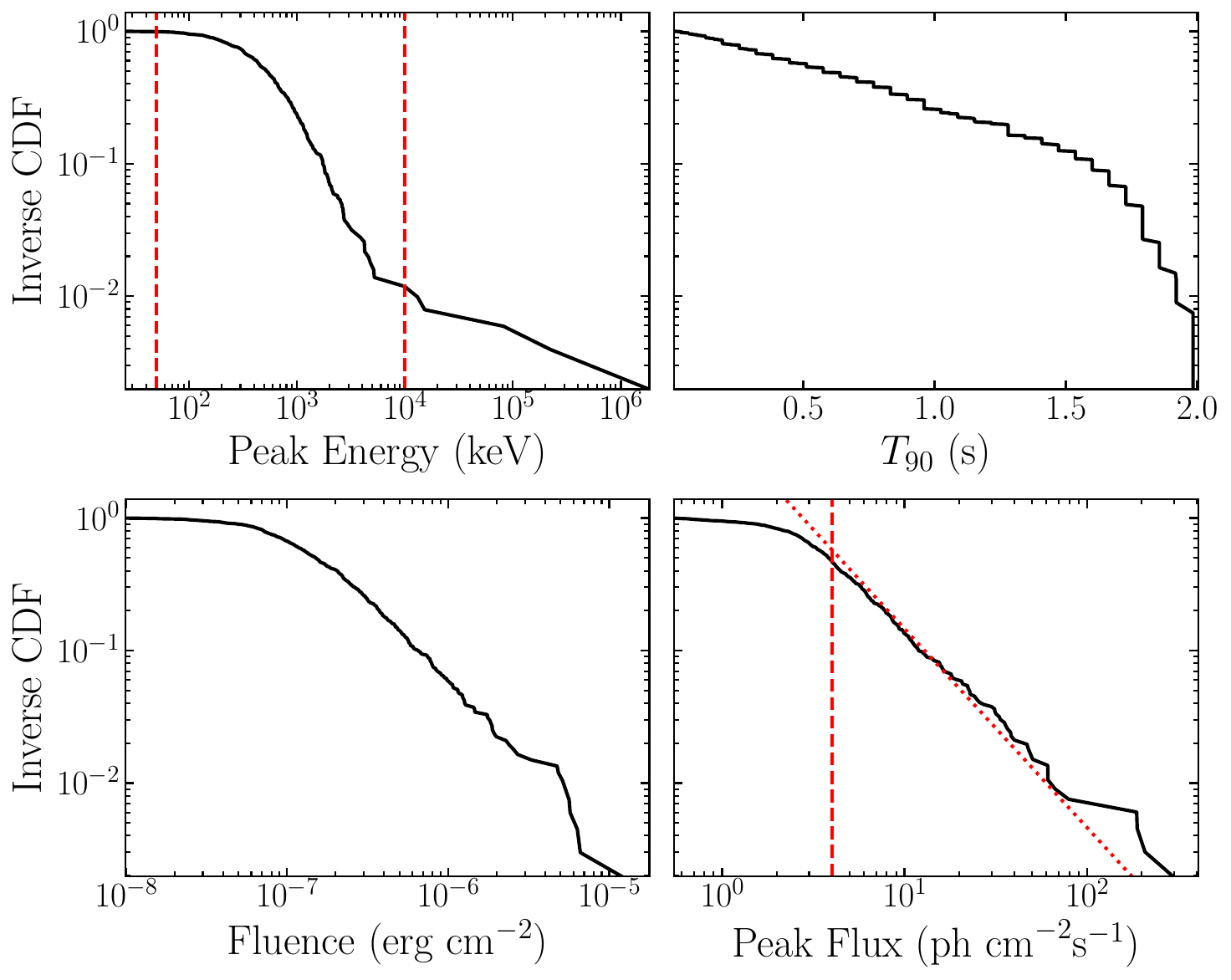}
    \caption{The inverse cumulative distributions of the observables retrieved from the \textit{Fermi-}GBM catalog. The additional quality cuts are shown with vertical red dashed lines. The dotted line shows a power law with index $-3/2$.}
    \label{fig:data_sel}
\end{figure}

We provide the distributions of the \textit{Fermi-}GBM data used in this work in \reffig{fig:data_sel}. As discussed in \refsec{sec:DATA} we have very similar cuts to previous works \citep{Salafia_2023, Ronchini2022}. The figure displays the complementary cumulative distribution functions for the four main observables which are peak energy, $T_{90}$, fluence, and peak flux. The vertical dashed lines indicate the specific cuts applied to define our sGRB sample ($F_p^{\text{lim}} \ge 4~\text{ph~cm}^{-2}~\text{s}^{-1}$, $T_{90} < 2$ s, and $E_p \in [50\text{ keV}, 10\text{ MeV}]$). We additionally plot a dotted power law line with index $-3/2$ which is the behavior expected form uniformly distributed GRBs \citep{vonKienlin2020}. This power law is used to define the peak flux cut to minimize selection effects.

\section{Description of Population Synthesis Models}
\label{app:pop_models}

In this work, we analyze the binary population synthesis models generated by \citet{Iorio2023} using the \textsc{sevn} code. These models explore the large parameter space of binary evolution uncertainties. While the original work provides a comprehensive technical description, we present here a simplified summary of the model variations most relevant to the formation of BNS mergers and their subsequent sGRB production. Table~\ref{tab:pop_descriptions} lists the model acronyms used throughout this paper and summarizes the major physical assumptions and differences among the populations.

\begin{table*}[h!]
\caption{Summary of BNS population synthesis models from \citet{Iorio2023}. We describe the primary physical assumption varied in each model and its observed impact on the merger rate.}
\label{tab:pop_descriptions}
\centering
\begin{tabular}{l p{3.5cm} p{9.5cm}}
\toprule
\textbf{Model} & \textbf{Physics Varied} & \textbf{Description \& Impact on Rates} \\
\midrule
F & Fiducial & The baseline model using standard assumptions for stellar winds and mass transfer, using a mass-dependent supernova kick model that allows for relatively high survival rates \citep{Giacobbo2020}. \\
\addlinespace
\multicolumn{3}{l}{\textit{Supernova Natal Kicks}} \\
\addlinespace
K265 & High Kicks & Neutron stars receive strong birth kicks (Maxwellian $\sigma = 265$ km/s). Significantly reduces the merger rate (by factor $\sim 4$ vs F) as binaries are disrupted. \\
K150 & Low Kicks & Neutron stars receive weaker birth kicks ($\sigma = 150$ km/s), increasing the rate compared to K265, but still yields fewer mergers than F. \\
\addlinespace
\multicolumn{3}{l}{\textit{Mass Transfer Stability}} \\
\addlinespace
QCBSE & Standard Stability & Uses restrictive criteria to decide if mass transfer is stable. Drastically suppresses the BNS rate (by $\sim 7$) as giant donors undergo unstable transfer and merge prematurely. \\
QCBB & Alternative Stability & 
Assumes mass transfer from pure-Helium stars is always stable. This avoids for some stars a final, often fatal Common Envelope phase where they would otherwise merge prematurely, allowing more systems to survive and 
eventually merge as BNSs.
\\
RBSE & Accretion Efficiency & Assumes the companion star accretes matter less efficiently during mass transfer, giving a moderate reduction in merger rates compared to F. \\
\addlinespace
\multicolumn{3}{l}{\textit{Common Envelope (CE) Physics}} \\
\addlinespace
LK, LC, LK & High Binding Energy & LK and LC assume tightly bound envelopes. This suppresses BNS formation as binaries fail to eject the envelope and merge as stars. Similarly LX uses the binding energy prescriptions from \citet{Xu2010b} which also reduces merger rates.\\
OPT & Optimistic CE & Allows Hertzsprung Gap donors to survive a CE phase. While it allows survival where standard models fail, this model is built upon the restrictive QCBSE physics. \\
\addlinespace
\multicolumn{3}{l}{\textit{Other Assumptions}} \\
\addlinespace
F19 & Pair Instability & Uses Pair Instability prescriptions from \citet{Farmer2019}. This mainly affects the black hole mass spectrum and has negligible impact on BNS merger rates. \\
F5M & Sampling Check & A high-resolution run of F. Used to verify that merger rates are not artifacts of statistical sampling noise. \\
SND & Supernova Engine & Uses a delayed supernova mechanism. This yield comparable total rates to F but alters the mass distribution of the compact objects by allowing some progenitors to not directly collapse into a black-hole. \\
QHE & Homogeneous Evolution & Allows rapidly rotating stars to mix chemically. Total merger rates are similar to F, though formation channels differ as these stars do not form Red Giants and remain compact. \\
NT & No Tides & Disables tidal forces. This results in a minor reduction in merger rates compared to F. \\
NTC & Circularization & Allows binaries to merger sooner during Roche lobe overflow. Increases the merger rate ($\sim 2$) by preventing collisions in eccentric orbits. \\
\bottomrule
\end{tabular}
\tablefoot{All models listed above are combined with four variations of the common envelope efficiency parameter, $\alpha_{\text{CE}} \in \{0.5, 1.0, 3.0, 5.0\}$, resulting in the 64 total populations analyzed in this work.}
\end{table*}

\section{MCMC convergence and posterior predictive checks}
\label{app:mcmc}
\begin{figure}[t!]
    \centering
    \includegraphics[width=\linewidth]{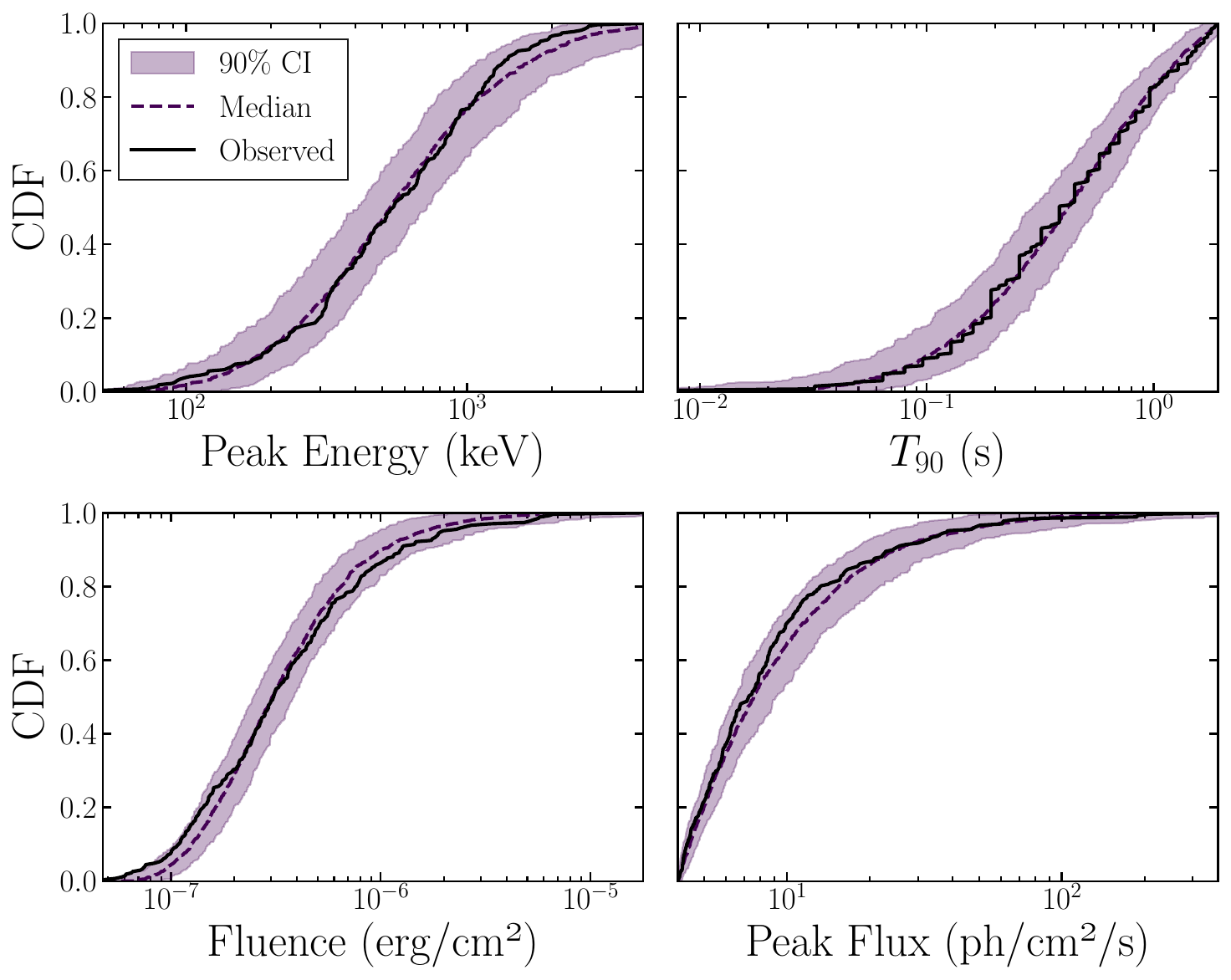}
    \caption{Cumulative distributions of the observables for real \textit{Fermi-}GBM sGRBs (solid line) and simulated sGRBs associated with our fiducial BNS population.}
    \label{fig:posterior_predictive}
\end{figure}

\begin{figure}[t!]
    \centering
    \includegraphics[width=\linewidth]{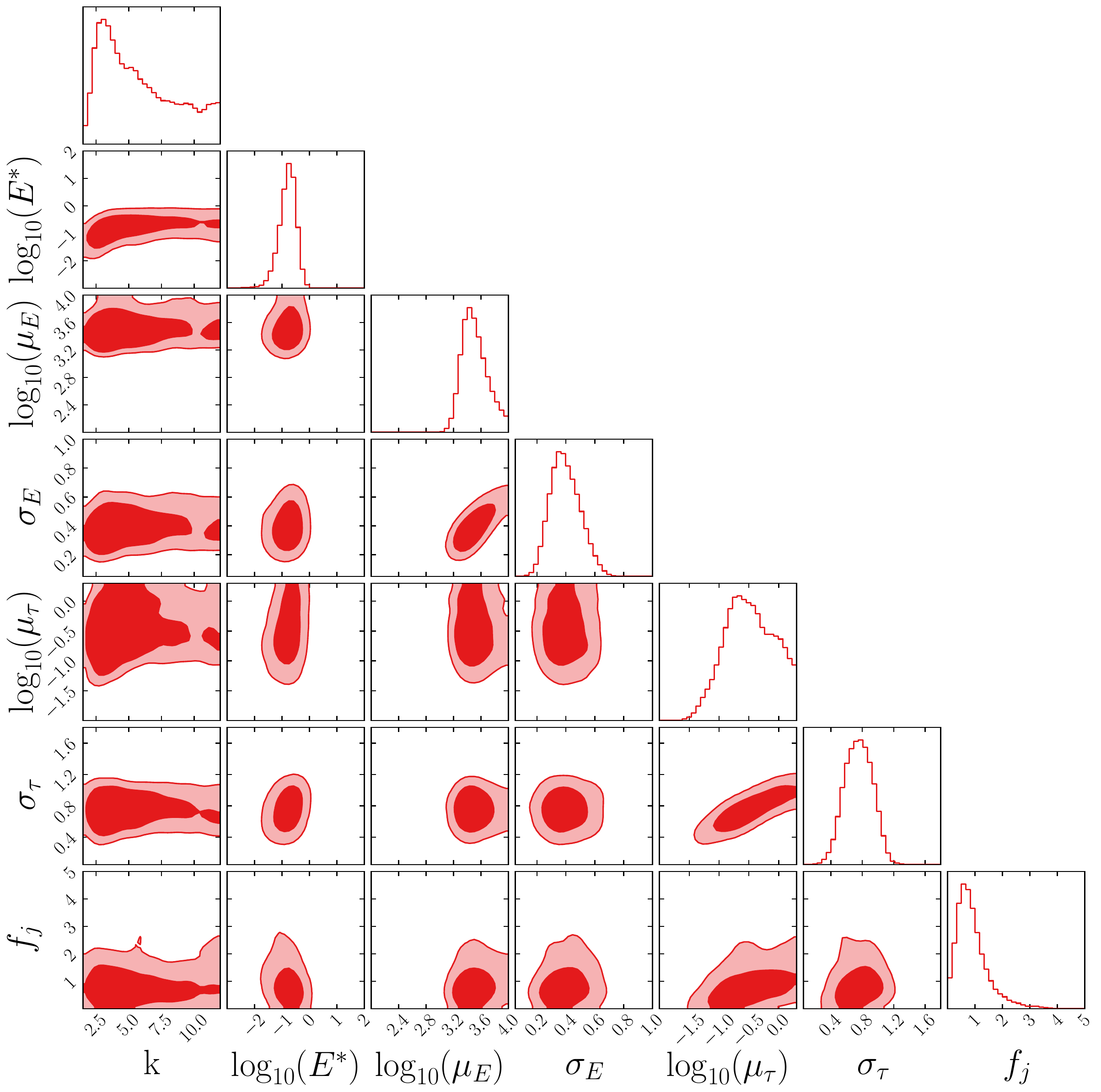}
    \caption{Posterior of the universal structured model for the fiducial population. The 1 and 2 $\sigma$ contours are shown.}
    \label{fig:corner_plot}
\end{figure}

\begin{figure}[t!]
    \centering
    \includegraphics[width=\linewidth]{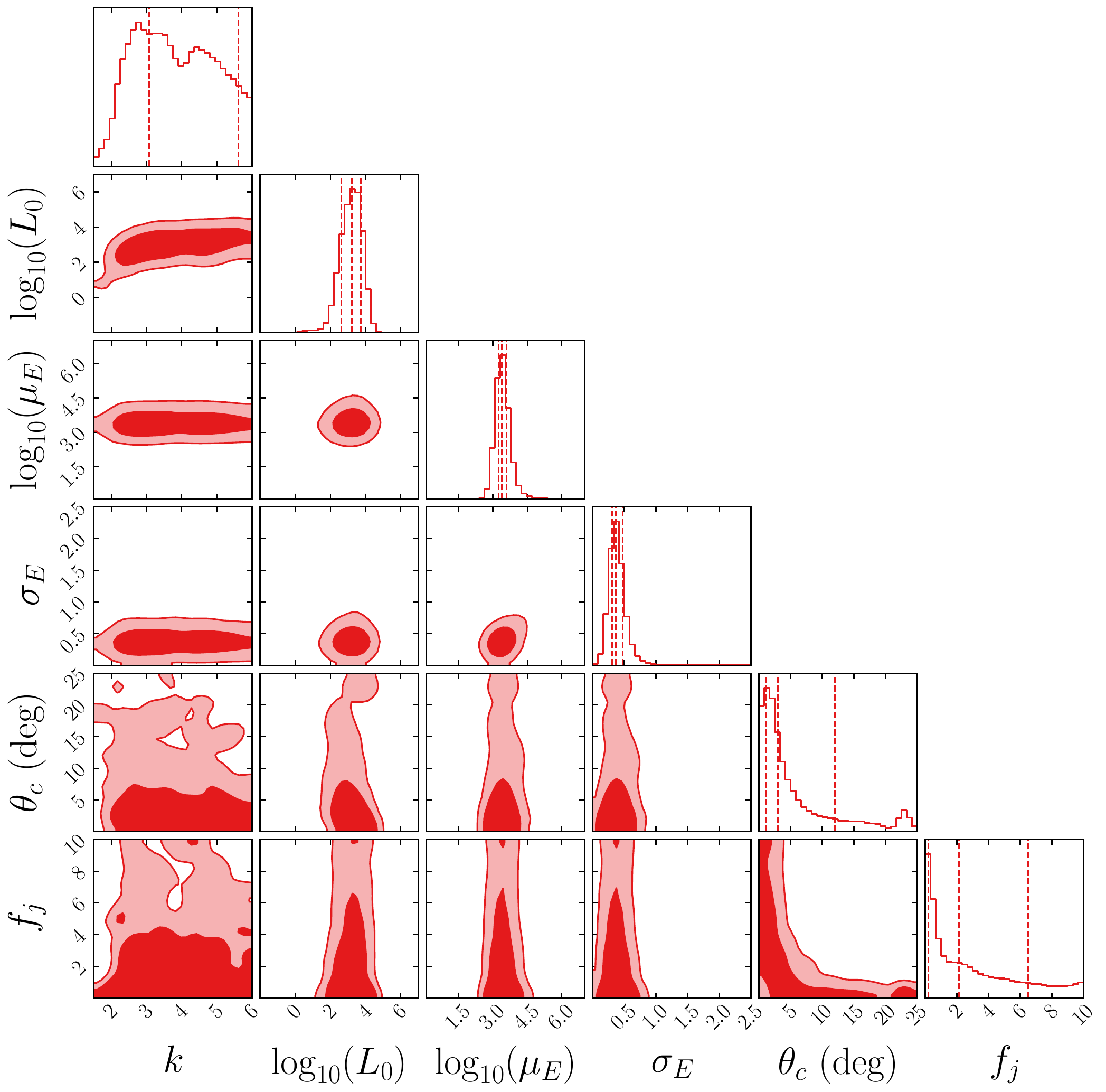}
    \caption{Posterior of the universal top-hat model for the fiducial population. The 1 and 2 $\sigma$ contours are shown.}
    \label{fig:simp_model_corner} 
\end{figure}

To ensure the MCMC simulations reached a converged posterior distribution, we monitored the integrated autocorrelation time, $\tau$, as defined in \citet{Foreman_Mackey_2013}. We require the total chain length to be at least 50 times greater than $\tau$, a criterion typically satisfied within 30,000 to 40,000 iterations, indicating well-sampled posteriors. With convergence established, we perform a posterior predictive check to assess the model’s goodness-of-fit. We generated synthetic sGRB catalogs by randomly sampling parameter sets from the converged chains and comparing the resulting Empirical Cumulative Distribution Functions (ECDFs) for the four key observables against the \textit{Fermi-}GBM data. We show in \reffig{fig:posterior_predictive} the simulated median and 90\% credible intervals having very good agreement with the observed distributions. The test is done on our Fiducial population. This confirms that the model successfully reproduces the fundamental features of the sGRB population. The resulting posterior probability distributions for the Universal Structured jet and the universal top-hat models are shown in \reffig{fig:corner_plot} and \reffig{fig:simp_model_corner}, with relative medians summarized in Table~\ref{tab:combined_posteriors}.
\begin{table}[h!]
    \centering
    \caption{Median posterior values and 90\% credible intervals for the parameters of the sGRB emission models using the fiducial BNS population.}
    \label{tab:combined_posteriors}
    \small
    \setlength{\tabcolsep}{3.5pt}
    \begin{tabular}{lc lc}
    \toprule
    \multicolumn{2}{c}{\textbf{Structured jet}} & \multicolumn{2}{c}{\textbf{Universal top-hat jet}} \\
    \cmidrule(r){1-2} \cmidrule(l){3-4}
    Parameter & Value & Parameter & Value  \\
    \midrule
    $k$ & $5.23_{-2.33}^{+4.16}$ & $k$ & $6.74_{-3.23}^{+3.55}$ \\
    $\log_{10}(E^*/10^{49}\,\text{erg})$ & $-0.75_{-0.31}^{+0.18}$ & $\log_{10}(L^*
/10^{49}\,\text{erg s}^{-1}
)$ & $3.23_{-0.57}^{+0.47}$ \\
    $\log_{10}(\mu_E/\text{keV})$ & $3.48_{-0.13}^{+0.22}$ & $\log_{10}(\mu_E/\text{keV})$ & $3.42_{-0.15}^{+0.37}$ \\
    $\sigma_E$ & $0.39_{-0.08}^{+0.11}$ & $\sigma_E$ & $0.39_{-0.08}^{+0.15}$ \\
    $\log_{10}(\mu_t/\text{s})$ & $-0.50_{-0.38}^{+0.48}$ & $\theta_c$ (deg) & $2.74_{-1.66}^{+4.38}$ \\
    $\sigma_t$ & $0.76_{-0.18}^{+0.17}$ &  \\
    $f_j$ & $0.78_{-0.36}^{+0.61}$ & $f_j$ & $2.87_{-2.36}^{+4.43}$ \\
    \bottomrule
    \end{tabular}
\end{table}

It is important to note that the resulting median characteristic energy posterior has a median at $E^*\sim10^{48}-10^{49}\,\text{erg}$. This value may appear low compared to the canonical beaming-corrected kinetic energies of sGRBs, which are often inferred to be in excess of $10^{50}\,\text{erg}$ \citep[e.g.][]{Berger2014}. This apparent discrepancy arises directly from our model's parameter definitions. Our $E^*$ represents the characteristic value of the radiated energy, distributed as a cut-off power-law, not the total beaming-corrected kinetic energy of the outflow. Taking into account this definition, our model produces a population of synthetic bursts whose observable fluences and fluxes are consistent with the distributions seen by \textit{Fermi-}GBM.


\section{Impact of Physical Priors on the Inferred Jet Fraction}
\label{app:boundary_effect}

In our primary analysis, we allow the jet fraction parameter $f_j$ to explore values well in excess of unity (setting the prior upper bound $f_{\text{max}} = 10$). While physically impossible, treating $f_j$ as an unbounded effective parameter is essential for quantifying the tension between BNS population models and sGRB observations.

\begin{figure*}[t]
    \centering
    \includegraphics[width=\textwidth]{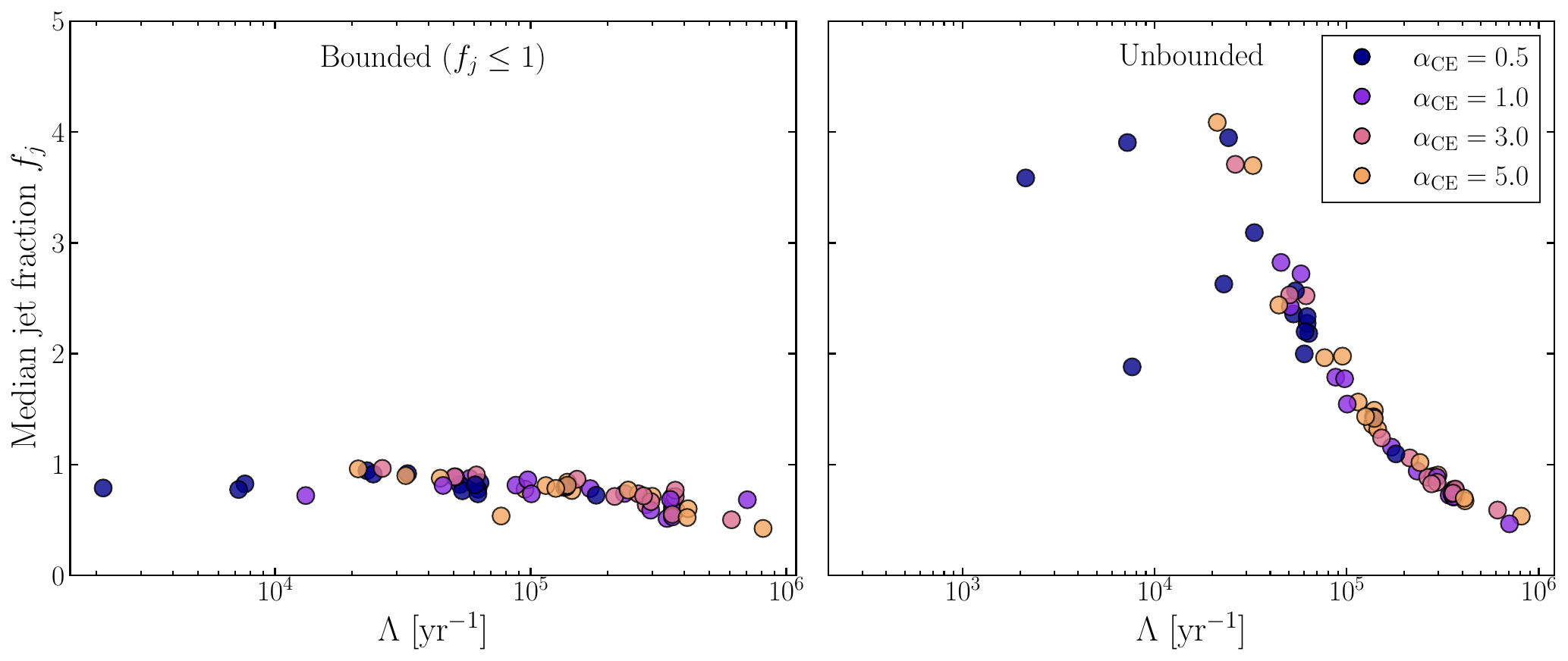}
    \caption{Comparison of the inferred median jet fraction $f_j$ as a function of total BNS rate $\Lambda$. The left panel shows bounded models, the right unbounded models. Zoomed in on the region $f_j \in [0, 5]$ for clarity. Both cases assume a universal structured jet.}
    \label{fig:boundary_saturation}
\end{figure*}

In order to see the effect of imposing a strict physical prior ($f_j \le 1$), we repeat the analysis with this constraint and we found that for models with low intrinsic merger rates ($R_{\text{BNS}}(0) \lesssim 50 \, \text{Gpc}^{-3}\text{yr}^{-1}$), the posterior $f_j$ saturates against the upper boundary of 1. In contrast, models with higher intrinsic rates find their optimal $f_j$ well within the physical boundary. \reffig{fig:boundary_saturation} illustrates this effect for the universal structured jet. With an unbounded prior, the median $f_j$ continuously rises as the BNS rate decreases. This allows us to distinguish between a model that requires $f_j \approx 1$ (corresponding to a physical scenario where all BNS mergers produce a jet) and one requiring $f_j \gg 1$ (which represents a fundamental breakdown of the sole BNS-progenitor hypothesis).

\section{Example narrowest geometry}
\label{app:narrowest}
\begin{figure}[t]
    \centering
    \includegraphics[width=\linewidth]{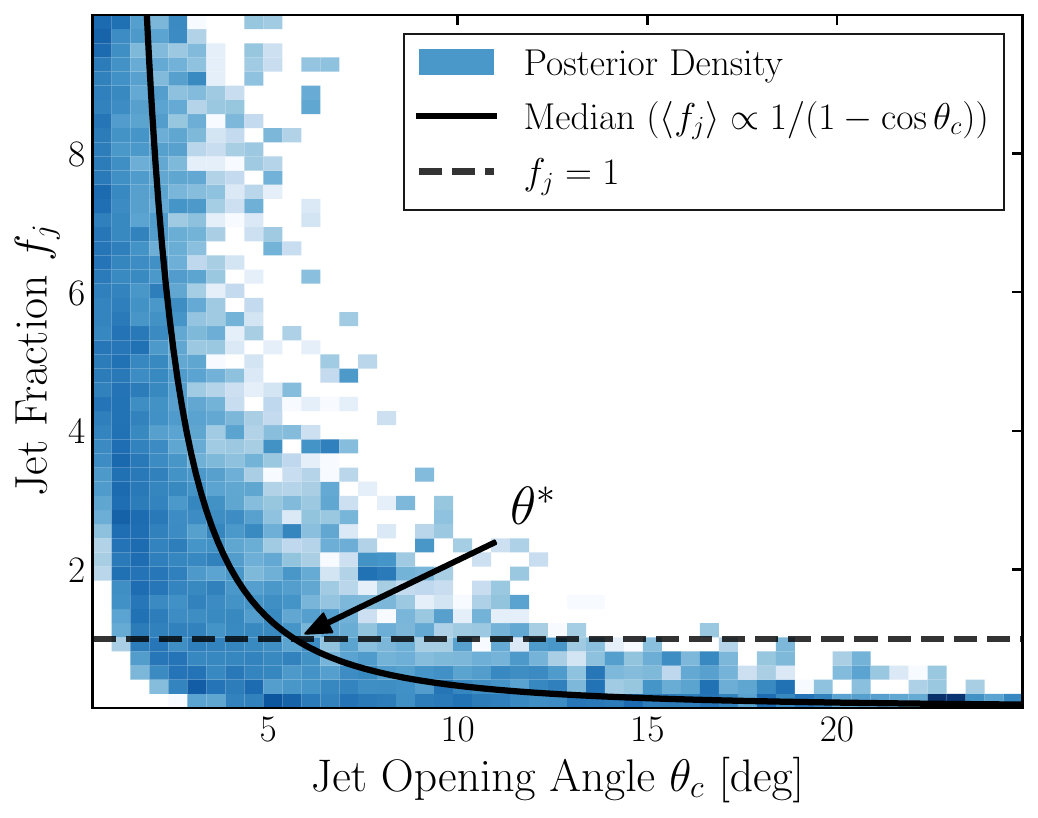} 
    \caption{Example narrowest geometry for the fiducial population using a universal top-hat jet. On top of the joint posterior $P(\theta_c, f_j)$ a solid line shows the median $f_j$ as a function of $\theta_c$. The intersection with $f_j = 1$ defines the minimum characteristic opening angle, $\theta^*$.}
    \label{fig:theta_c_threshold_visual}
\end{figure}

We give a visual example on how to recover the narrowest geometry for a universal top-hat model in \ref{fig:theta_c_threshold_visual}. Given that for all models the product between jet fraction $f_j$ and beaming factor $f_b$ is monotone as a function of the structure parameter ($\theta_c$, $\theta_c^\text{max}$ and $\theta_c^\text{med}$) we can find the minimum value of these parameters such as to have a physical jet fraction.

\section{Summary of Model Viability Criteria}  
\label{app:table_details}

In this appendix, we detail the criteria used to construct the summary of results presented in Table~\ref{tab:viability_grid}. The 64 BNS population models, characterized by different binary evolution prescriptions and common envelope efficiencies ($\alpha_{\text{CE}}$), are evaluated based on two primary constraints. \textit{Physical viability:} A model is marked as physical (\ch) if the median of its inferred jet fraction posterior distribution lies within the physical regime (median $f_j \leq 1$), which is equivalent to $P(f_j \leq 1) \geq 0.5$. If the median $f_j > 1$, the model is marked as non-physical ($\x$). For models with non-universal structure we fix the median $\theta_c$ at the median value of \citet{Rouco_2023} ($\sim6^\circ$). \textit{Geometric Consistency:} For models where the jet structure parameters are free (universal top-hat, non-universal flat, and non-universal log-normal), we assess if the geometry required to maintain a physical $f_j$ is consistent with afterglow observations. We use the 90\% credible interval upper limit from \citet{Rouco_2023} of $15.4^\circ$.
\begin{itemize}
    \item For the universal top-hat and flat models, the index $^a$ is applied if the minimum characteristic opening angle $\theta^*$ required to satisfy median $f_j \leq 1$ exceeds $15.4^\circ$.
    \item For the log-normal model, the index $^b$ is applied if more than 10\% of the inferred jet population is required to have an opening angle $\theta_c > 15^\circ$ to satisfy the rate requirement. 
\end{itemize}
The combination of these marks allows for a rapid identification of viable BNS populations which are those that provide enough progenitors to match the sGRB rate while maintaining jet geometries consistent with observed afterglows.

\begin{table*}[t]
\centering
\caption{Physical viability and geometric consistency of BNS population models. Checkmarks (\ch) indicate physical jet fractions (with median $f_j \leq 1$), while crosses ($\x$) indicate non-physical scenarios (median $f_j > 1$). For the universal top-hat, we fix the aperture angle at $\theta_c = 6.1\,\text{deg}$ which is the median value cited in \citealt{Rouco_2023}. Analogously, for both Flat and Log-Normal structures, we fix the median of the $\theta_c$ distribution at $6.1\,\text{deg}$ .}
\label{tab:viability_grid}
\small

\begin{tabular}{l | cccc | cccc | cccc | cccc}
\toprule
 & \multicolumn{4}{c|}{\textbf{Universal top-hat}} & \multicolumn{4}{c|}{\textbf{Non-univ. flat}} & \multicolumn{4}{c|}{\textbf{Non-univ. log-normal}} & \multicolumn{4}{c}{\textbf{Universal structured}} \\
$\alpha_{\text{CE}}$ & 0.5 & 1.0 & 3.0 & 5.0 & 0.5 & 1.0 & 3.0 & 5.0 & 0.5 & 1.0 & 3.0 & 5.0 & 0.5 & 1.0 & 3.0 & 5.0 \\
\midrule
F     & $\x^a$        & \ch       & \ch & $\x$   & $\x$   & \ch & \ch & $\x$   & \ch & \ch & \ch & \ch & $\x$   & \ch & \ch & $\x$   \\
K265  & $\x^a$      & $\x$      & $\x$   & $\x^a$ & $\x$   & $\x$   & $\x$   & $\x$   & \ch$^b$ & \ch & \ch & \ch$^b$ & $\x$   & $\x$   & $\x$   & $\x$   \\
K150  & $\x$        & $\x$      & \ch & $\x$   & $\x$   & $\x$   & $\x$   & $\x$   & \ch & \ch & \ch & \ch & $\x$   & $\x$   & $\x$   & $\x$   \\
QCBSE & $\x^{a,c}$  & $\x$      & $\x^a$ & $\x^a$ & $\x$   & $\x$   & $\x$   & $\x$   & $\x^b$ & \ch & \ch$^b$ & \ch$^b$ & $\x$   & $\x$   & $\x$   & $\x$   \\
QCBB  & \ch         & $\x$      & $\x$   & $\x$   & \ch & \ch$^a$ & $\x$   & $\x$   & \ch & \ch & \ch & \ch$^b$ & $\x$   & $\x$   & $\x$   & $\x$   \\
RBSE  & $\x$        & $\x^a$       & \ch & $\x$   & $\x^a$ & \ch & \ch & $\x^a$ & \ch$^b$ & \ch & \ch & \ch$^b$ & $\x$   & \ch & \ch & $\x$   \\
LK    & $\x^{a, c}$  & $\x^c$     & $\x$   & \ch & $\x$   & $\x$   & \ch & \ch & $\x^b$ & \ch$^b$ & \ch & \ch & $\x$ & $\x$   & $\x$   & \ch \\
LC    & $\x^a$      & $\x$      & \ch & \ch & $\x$   & $\x$   & \ch & \ch & \ch$^b$ & \ch & \ch & \ch & $\x$   & $\x$   & \ch & \ch \\
LX    & $\x^{a,c}$  & $\x$      & \ch & \ch & $\x$   & $\x$   & \ch & \ch & $\x^b$ & \ch$^b$ & \ch & \ch & $\x$   & $\x$   & \ch & \ch \\
OPT   & $\x^{a,c}$  & $\x$      & \ch & $\x^c$ & $\x$   & $\x$   & \ch & \ch & $\x^b$ & \ch & \ch & \ch & $\x$   & $\x$   & \ch & \ch \\
F19   & $\x^a$        & \ch       & \ch & $\x$   & $\x^a$ & \ch & \ch & $\x$   & \ch & \ch & \ch & \ch & $\x$   & \ch & \ch & $\x$   \\
F5M   & $\x^a$        & \ch       & \ch & $\x^a$   & $\x$   & \ch & \ch & $\x$   & \ch & \ch & \ch & \ch & $\x$   & \ch & \ch & $\x$   \\
SND   & $\x$        & $\x^c$       & \ch & $\x$   & $\x^a$ & \ch & \ch & $\x$   & \ch & \ch & \ch & \ch & $\x$   & \ch & \ch & $\x$   \\
QHE   & $\x$        & \ch       & \ch & $\x$   & $\x$   & \ch & \ch & $\x$   & \ch & \ch & \ch & \ch & $\x$   & \ch & \ch & $\x$   \\
NT    & $\x$        & $\x^c$       & $\x^c$ & $\x$   & $\x$   & \ch & \ch & $\x$   & \ch & \ch & \ch & \ch & $\x$   & \ch & \ch & $\x$   \\
NTC   & $\x$        & \ch       & \ch & \ch & $\x$   & \ch & \ch & \ch & \ch & \ch & \ch & \ch & $\x$   & \ch & \ch & $\x$ \\
\bottomrule
\end{tabular}

\begin{flushleft}
$^a$ Required minimum characteristic opening angle $\theta^* > 15.4^\circ$ (90\% upper limit from \citealt{Rouco_2023}).\\
$^b$ More than 10\% of the inferred jet population exceeds $\theta_c = 15^\circ$.\\
$^c$ The required $\theta^*$ was greater than our defined prior bounds ($< 25^\circ$) and could not be explicitly computed. 
\end{flushleft}
\end{table*}

\section{Redshift distribution of the average time delay}
\label{app:tdvsz}
\begin{figure*}[t]
    \centering
    \includegraphics[width=\textwidth]{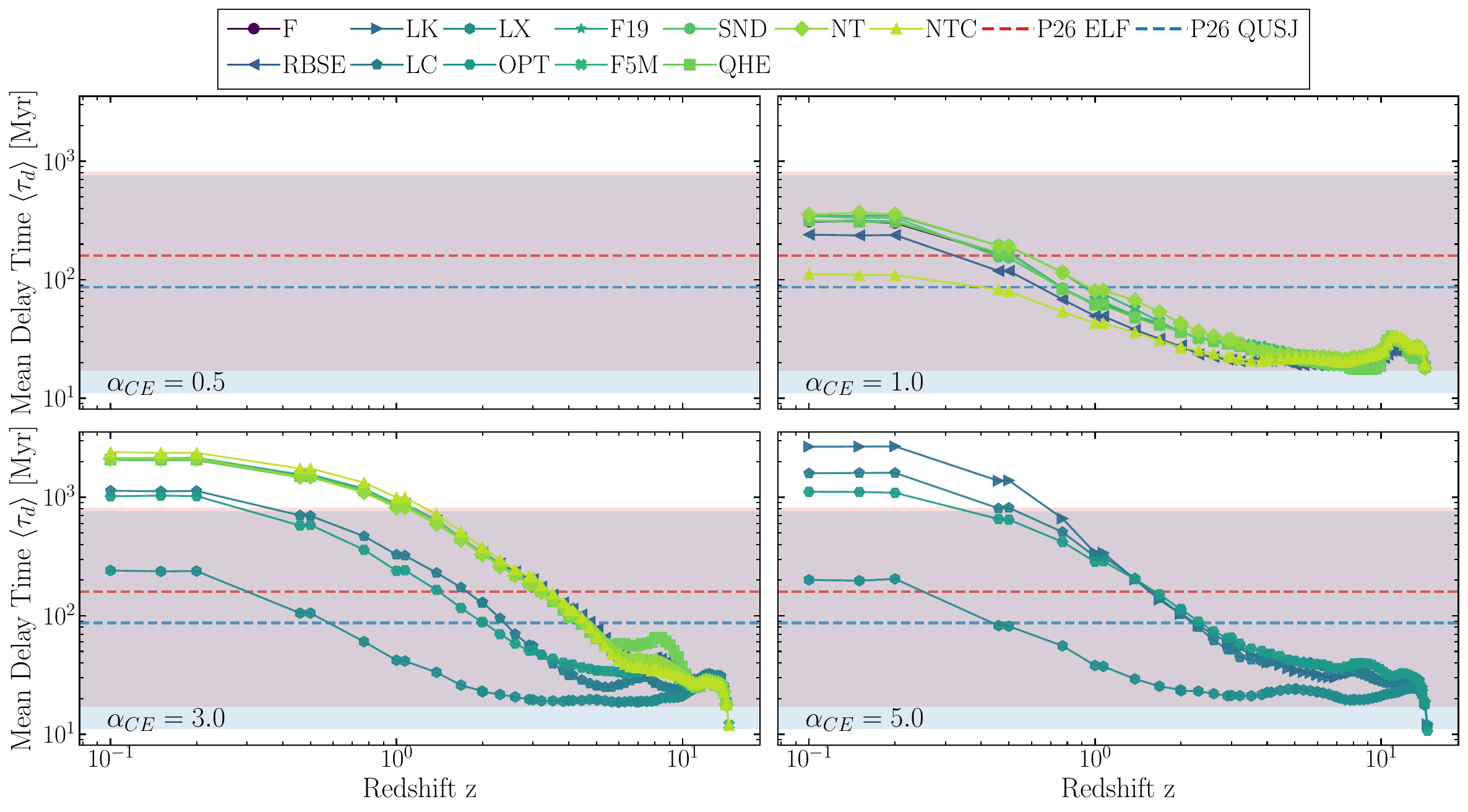} 
    \caption{Average delay time $\langle \tau_d \rangle$ evolution with redshift for physical population (median $f_j \leq 1$) assuming the universal structured jet compared to the median and 90\% credible intervals inferred in \citet{pracchia2026shortgammarayburstprogenitors}. Plots with erratic spikes have a very small number of mergers. See Table~\ref{tab:pop_descriptions} for a description of each model variation (top right of each plot). See \reffig{fig:tdvsz_np} for non physical populations. For consistency with \reffig{fig:tdvsz_np} we keep the panel for $\alpha_{\text{CE}}=0.5$ even though no physical models exist for that value in the universal structured jet case.}
    \label{fig:tdvsz_phys}
\end{figure*}

\begin{figure*}[t]
    \centering
    \includegraphics[width=\textwidth]{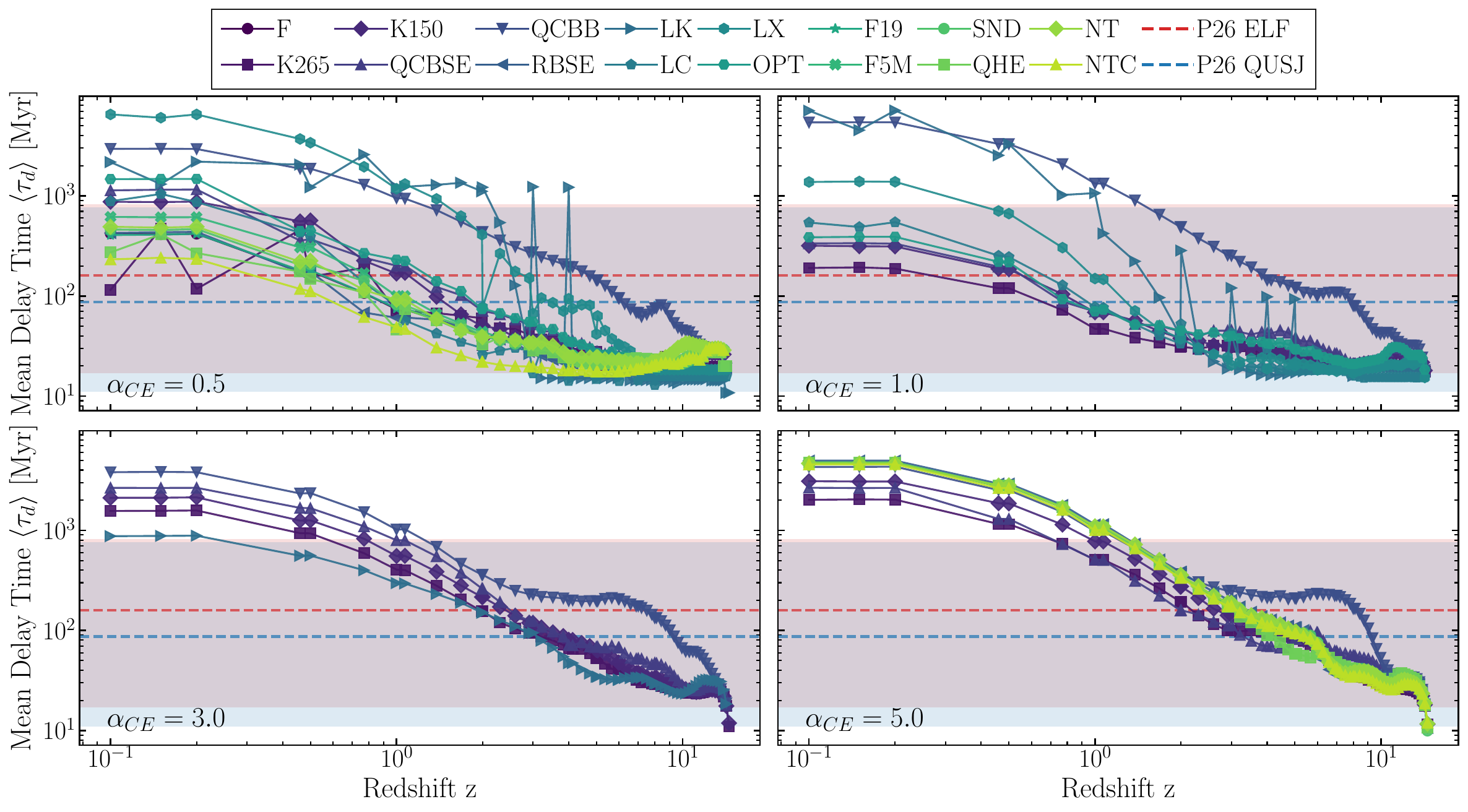} 
    \caption{ Same as \reffig{fig:tdvsz_phys} for non physical populations (median $f_j > 1$).}
    \label{fig:tdvsz_np}
\end{figure*}

In \reffig{fig:tdvsz_phys} (for physical models) and \reffig{fig:tdvsz_np} (for the remaining models), we show the redshift evolution of the average delay time, $\langle \tau_d \rangle$, for each of our 64 models. Across all models, we observe that the delay times decrease by approximately two orders of magnitude as redshift increases. While the physical models exhibit the best agreement with the population averages derived by \citet{pracchia2026shortgammarayburstprogenitors}, it is crucial to note that the average delay time is sensitive to the distribution tail. Specifically, $\langle \tau_d \rangle$ can be skewed by the large delay times found at low redshifts ($z < 1$). Since this redshift range corresponds to the volume where sGRBs are most likely to be detected, the observed population is inherently biased towards these longer delay times, this underscores the importance of accounting for selection effects when constructing sGRB catalogs from fixed population synthesis models. A similar conclusion is reached by \citet{pracchia2026shortgammarayburstprogenitors}, who note that selection effects, specifically when applying a peak flux cut that is too low, can bias the sample towards larger delay times.

\end{document}